\documentclass[journal]{IEEEtran}
\usepackage{verbatim}
\usepackage{cite}
\usepackage{graphicx}
\usepackage{times}
\usepackage{latexsym}

\usepackage[mathscr]{eucal}
\usepackage{array}
\usepackage{fancyhdr}
\usepackage{amsmath,amssymb}
\usepackage{bm} 
\usepackage{subfigure}
\usepackage{multirow}

\usepackage{algpseudocode}
\usepackage{algorithm}
\usepackage{amsfonts}

\ifCLASSINFOpdf
\else
\fi

\newtheorem{thm}{Theorem}

\newtheorem{prop}{Proposition}

\newtheorem{rem}{Remark}

\newcommand{\qed}{\hfill\ensuremath{\blacksquare}}

\newcommand{\nn}{\nonumber}

\newcommand{\vect}[1]{{\lowercase{\mathbf{#1}}}}
\newcommand{\mat}[1]{{\uppercase{\mathbf{#1}}}}

\newcommand{\tr}{{\rm{tr}}}

\newcommand{\st}{{\rm{\mbox{s.t. }}}}
\newcommand{\tvec}{{\rm{vec}}}

\renewcommand{\a}{\vect{a}} 
\newcommand{\e}{\vect{e}}

\newcommand{\z}{\vect{z}}

\newcommand{\A}{\mat{A}}

\newcommand{\C}{\mat{C}}

\newcommand{\G}{\mat{G}}
\renewcommand{\H}{\mat{H}} 
\newcommand{\I}{\mat{I}}

\newcommand{\N}{\mat{N}}
\renewcommand{\P}{\mat{P}}

\newcommand{\R}{\mat{R}}
\renewcommand{\S}{\mat{S}}

\newcommand{\U}{\mat{U}}
\newcommand{\V}{\mat{V}}

\newcommand{\Y}{\mat{Y}}
\newcommand{\Z}{\mat{Z}}

\newcommand{\Rc}{{\cal R}}

\newcommand{\Ct}{{\tilde \C}}

\newcommand{\Ht}{{\tilde \H}}

\newcommand{\Pt}{{\tilde \P}}

\newcommand{\Rt}{{\tilde \R}}

\newcommand{\Cb}{{\mathbb C}}

\newcommand{\Eb}{{\mathbb E}}

\newcommand{\Lambdam}{\hbox{\boldmath$\Lambda$}}

\usepackage{xcolor}

\makeatother

\allowdisplaybreaks

\graphicspath{{Figure/}}

\begin{document}
\title{{Sum Secret Key Rate Maximization for TDD Multi-User Massive MIMO Wireless Networks}}

\author{
Guyue~Li,~\IEEEmembership{Member,~IEEE},
Chen~Sun,~\IEEEmembership{Member,~IEEE},
Eduard~Jorswieck,~\IEEEmembership{Fellow,~IEEE},
Junqing~Zhang, Aiqun~Hu,~\IEEEmembership{Member,~IEEE},
and You~Chen
\thanks{G. Li and Y. Chen are with the School of Cyber Science and Engineering, Southeast University, Nanjing, China. (e-mail: guyuelee@seu.edu.cn.)}
\thanks{C. Sun and A. Hu are with the National Mobile Communications Research Laboratory,
Southeast University, Nanjing, 210096, China. (e-mail: sunchen@seu.edu.cn, aqhu@seu.edu.cn.)}
\thanks{G. Li, C. Sun and A. Hu are also with the Purple Mountain Laboratories for Network and Communication Security, Nanjing, 210096, China}
\thanks{E. Jorswieck is with the Institute for Communications Technology, Technische Universität Braunschweig,  Germany. (e-mail: Jorswieck@ifn.ing.tu-bs.de)}
\thanks{J. Zhang is with the Department of Electrical Engineering and Electronics, University of Liverpool, Liverpool, L69 3GJ, United Kingdom. (email: junqing.zhang@liverpool.ac.uk.)}
\thanks{C. Sun is the corresponding author.}
}

\maketitle

\begin{abstract}
Physical-layer key generation (PKG) based on channel reciprocity has recently emerged as a new technique to establish secret keys between devices. Most works focus on pairwise communication scenarios with single or small-scale antennas. However, the fifth generation (5G) wireless communications employ massive multiple-input multiple-output (MIMO) to support multiple users simultaneously, bringing serious overhead of reciprocal channel acquisition. This paper presents a multi-user secret key generation in massive MIMO wireless networks. We provide a beam domain channel model, in which different elements represent the channel gains from different transmit directions to different receive directions. Based on this channel model, we analyze the secret key rate and derive a closed-form expression under independent channel conditions. To maximize the  {sum} secret key rate, we provide the optimal conditions for the Kronecker product of the precoding and receiving matrices and  propose an algorithm to generate these matrices with pilot reuse. The proposed optimization design can significantly reduce the pilot overhead of the reciprocal channel state information acquisition. Furthermore, we analyze the security under the channel correlation between user terminals (UTs), and propose a low overhead multi-user secret key generation with non-overlapping beams between UTs. Simulation results demonstrate the near-optimal performance of the proposed precoding and receiving matrices design and the advantages of the non-overlapping beam allocation.
\end{abstract}


\section{Introduction} 
\label{sec:introduction}
Physical-layer key generation (PKG) has emerged as a promising technique to share the symmetric key for cryptographic applications~\cite{zhang2016review}. Based on the reciprocity of the uplink and downlink channels, the communication ends, named Alice and Bob, can establish a pair of common channel information. When the channel information is converted into symmetric keys, Alice and Bob can use them for cryptography to safeguard data communication.
The secret keys can be regularly updated, since the channel information varies randomly over time. Furthermore, due to channel decorrelation effect~\cite{5492690}, any eavesdropper, named Eve, located half a wavelength away or more from Alice and Bob, will observe an uncorrelated channel information~\cite{Li2018High}. Thus, Eve cannot infer the secret key based on her own channel observations.

The key distribution is usually handled by the traditional public key cryptography techniques, which however have been facing a number of challenges to be applied in the future networks.
Firstly, public key needs to be distributed to different devices in advance, and the key distribution for massive devices is complicated~\cite{Zhang2017Securing}. Secondly, the distributed key for each device usually does not update for a long time which may incur security issues.
Thirdly, the public key cryptography may be compromised by the emerging quantum computers in the future~\cite{Chi2017Securing}.
Key generation can thus be a good alternative to complement when the public key cryptography is not suitable.


The PKG process generally contains four stages, namely channel probing, quantization, information reconciliation, and privacy amplification.
At the beginning, Alice and Bob alternately transmit pilot signals to obtain correlated channel measurements.
Then, they quantize these analog channel measurements into digital bits. Although the uplink and downlink channels are reciprocal, due to the calibration errors in uplink/downlink RF chains, the temporal variation of the channel and the noise, the measurements of the uplink and downlink channel are not identical but highly correlated.
Next, information reconciliation is used to enable Alice and Bob to agree on the same key through error detection protocols or error correction codes~\cite{Huth2016}.
Finally, privacy amplification eliminates any potential information leakage to eavesdroppers.

Based on channel reciprocity, the channel probing stage shares the common random sources between legitimate users to generate the secret keys. Most PKG implementations are realized in the time division duplex (TDD) mode in order to utilize the channel reciprocity.
Specifically, Alice and Bob alternately transmit the probing signals to estimate the channel state information (CSI), where the sampling time difference between them should be smaller than the channel coherence time, indicating that the coherence time limits the pilot overhead.
As the pilot overhead scales linearly with the number of antennas,
single antenna or small-scale multiple-input multiple-output (MIMO) communication systems, which is considered in most of the existing PKG schemes~\cite{Jorswieck2014Secret}, have enough time and frequency resources to obtain the highly correlated CSI for pairwise communication scenarios.

The fifth generation (5G) and beyond communication systems employ massive MIMO technologies to support extremely high throughput and multiple users~\cite{8241348}.
However, it is challenging to apply PKG with massive MIMO~\cite{Li2019Physical}.
As the number of antennas is extremely large in massive MIMO systems, it is impractical for the base station (BS) and the user terminal (UT) to estimate the instantaneous uplink and downlink channel information within the coherence time.
In some quasi-static scenarios, the coherence time may be long enough for pairwise channel estimations. However, the channel varies slowly such that the adjacent measurements are highly correlated, which will introduce redundancy and may finally result in failure of key generation. In the previous work, the self-correlation is eliminated by introducing signal preprocessing procedure after channel sounding, such as principal component analysis (PCA)~\cite{Li2018High}. This procedure also introduces great complexity due to the large data dimension.



Massive MIMO exploits spatial diversity and spatial signatures by allocating different beams/angles of transmitted signal to different directions of users, which enables multi-user communications. Key generation usually occurs between a pair of users. Exploiting massive MIMO for multi-user key generation will not be a straightforward extension from the pairwise key generation. This exploration is currently missing but very important as multi-user secret communications are very demanding.

This paper aims to address the above challenges by generating secret key with multi-users simultaneously in a narrow band massive MIMO system.
In particular, we first state the problems for intuitively expanding existing key generation schemes in a multi-user massive MIMO scenario and then propose a
new channel dimensionality reduction (CDR) based key generation scheme to address these problems. The main contributions of this paper are listed as follows:
\begin {itemize}
\item  We propose a novel CDR-based key generation scheme exploiting sparse property of the beam domain channel model. Legitimate users only need to estimate the effective channels at a few dominate beams, therefore the pilots lengths and channel auto-correlations are largely reduced. Furthermore, we derive a closed-form expression of the secret key rate, considering other UTs as non-colluding curious users.
\item We present an optimization approach to realize the maximal  {sum} secret key rate under the pilot reuse case where different UTs transmit the identical pilot signals.
Specifically, we design the precoding and receiving matrices to reduce the inter-user interference for muti-user communications.
The proposed approach reduces the pilot overhead that scales with the number of UTs.
\item   We provide a security analysis considering the spatial channel correlations between UTs, and reveal that the channel information on the overlapping beams may cause serious information leakage and provide little secret key rate.
Therefore, we propose a holistic multi-user secret key generation scheme, where the BS allocates non-overlapping beams to different UTs and multiple UTs can simultaneously generate secret keys with the BS using non-overlapping beams.
Numerical results demonstrate the performance improvement of our proposed multi-user secret key generation scheme.
\end {itemize}

The material in this paper has been partially accepted by IEEE ICC 2020. In our previous work, we focus on the pilot reuse case, and proposed a beam-domain secret key generation approach which can reduce the channel dimension and the pilot overhead in a multi-user massive MIMO system.
In this paper, we consider a general multi-user secret key generation framework and provide a general secret key rate, containing both the orthogonal pilot and the reused pilot cases. We prove the optimality of our proposed algorithm in the reused pilot case. Furthermore, we add a security analysis by considering the channel correlation and conduct that the BS employing non-overlapping beams to generate secret keys with different UTs.

We use the following notation throughout the paper:
Upper (lower) bold-face letters denote matrices (column vectors); ${\bf {I}}$ denotes the identity matrix while its subscript, if needed,  represents its dimensionality.
Let $\e_i = [0,0,\cdots,0,1,0,\cdots,0]$ denote a unit vector with the $i$th unit element and $\lambda_i(\A)$ represent the $i$th sorted eigenvalue of matrix $\A$.
The superscripts $(\cdot)^H, (\cdot)^T, (\cdot)^*$ stand for the conjugate-transpose, transpose, and conjugate of a matrix, respectively.
We use $\mathbb{E}\{\cdot\}$ to denote ensemble expectation and $\tr(\cdot), \det(\cdot)$ to represent matrix trace and determinant operations, respectively.
The $\tvec(\cdot)$ operator stacks the columns of a matrix into a tall vector, and the symbol $\otimes$ denotes the Kronecker product of two matrices.
The inequality $\A\succeq {\bf 0}$ means that $\A$ is Hermitian positive semi-definite.
We use $[{\bf A}]_{m,n}$ to denote the $(m,n)$-th element of matrix $\bf A$.
The covariance matrix of combined random vectors is defined as
$\Rc_{\z_1\z_2\cdots\z_{N_z}} = \Eb\{ \z\z^H \}$, where
$\z = \left[\z_1^T, \z_2^T, \cdots, \z_{N_z}^T  \right]^T$.

\section{Related Work}
There are very few papers working on key generation with massive MIMO.
The work in~\cite{Jiao2} employed new channel characteristics, e.g., virtual angle of arrival (AoA) and angle of departure (AoD), to generate a shared secret key between two devices.
Furthermore, Jiao~\textit{et al} added a small perturbation angle into the AoA of the transmitter as the common randomness to improve the secret key rate constrained  by low dynamic of the channel~\cite{Jiao1}.
However, these works only focus on the AoA and AoD to generate the secret key, while the optimal design maximizing the secret key rate is missing.

While pairwise key generation has been extensively investigated, group and multi-user key generation yet receives less attention~\cite{zhang2016review}.
{ Note that, although both schemes have multiple users participating in the key generation process, we distinguish them in this paper.
In the group key generation, all of the users share a common secret key.
The multi-user key generation studied in this paper, refers to a particular case that that each BS-UT pair has a different secret key.}
Among the existing group key generation protocols~\cite{liu2014group,thai2019secret}, the majority of them still perform channel probing in a pairwise manner, resulting in an extremely large overhead and low efficiency.
Hence, those works related to PKG among multiple nodes through the optimization of probing rates at individual node pair and channel probing schedule do not scale in this context~\cite{jin2018physical}.
In the multi-user key generation, Zhang~\textit{et al} cleverly exploited the multi-user mechanism of OFDMA modulation by assigning non-overlapping subcarriers to different users~\cite{zhang2019design}. However, there is no work exploiting the spatial diversity of massive MIMO to enable multi-user key generation.

Therefore, there is a clear need to investigate key generation with massive MIMO with special consideration to the multi-user applications, in both theoretical analysis on the secret key rate and the design of practical protocols.

\section{System Model and Problem Statement} 
\label{sec:system_model}
\begin{figure}[!t]
\centering
{\includegraphics[width=0.48\textwidth]{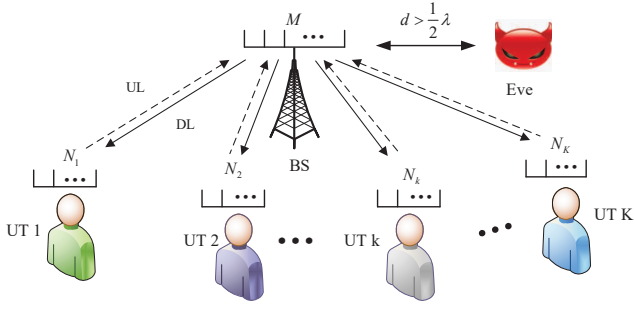}}
\caption{System model of multi-user secret key generation.}
\label{fig:system}
\end{figure}

This paper considers a narrow-band star topology network, where a BS simultaneously generates secret keys $\kappa = \{\kappa_1, \kappa_2, \cdots, \kappa_K\}$ with $K$ UTs, as shown in Fig.~\ref{fig:system}.
The BS is equipped with $M$ antennas and the $k$th UT is equipped with $N_k$ antennas. In the 5G and beyond wireless communications, massive MIMO is the key technology to increase the transmission rate.
Under the TDD operation, based on the reciprocal uplink and downlink channels, the BS generates the pairwise key $\kappa_k$ with the $k$th UT.


We consider passive eavesdropping here and the active attacks are out of scope in this paper. Specifically, an eavesdropper  wants to eavesdrop the secret key in the set of $\kappa$ based on her own channel observations and all the information exchanged over the public channel.
As a general assumption in PKG,
{we assume that Eve is located at least half wavelength away from BS and all of the UTs, because half wavelength is very close, e.g., 6.25 cm for 2.4GHz and Eve might get detected within a distance below that.}
Therefore, Eve's channel observations are assumed to be independent of that between the BS and UTs.
Besides of Eve, we also consider the potential unintended hearing from other UTs.
When two UTs are located close to each other, they may have correlated channel observations.
These UTs are treated as curious users, i.e.,  each of them does not intend to eavesdrop keys of other users.
They also do not collude with other UTs or Eve.

The secret keys are extracted from the reciprocal wireless channel information. Thus, we will first introduce the channel model and then state the problems and challenges in multi-user massive MIMO key generation.

\subsection{Channel Model} 
\label{sub:channel_model}

We consider a geometric channel model with $N_P$ paths.
Then, the $N_k\times M$ physical MIMO channel matrix in the downlink
associated with the $p$th path of the $k$th UT can be expressed as \cite{DTse}
\begin{align}\label{eq:1}
    \H_{k,p}^{DL} = \alpha_{k,p} \a_{UT,k}(\theta_{k,p}) \a^H_{BS}(\varphi_{k,p}),
\end{align}
where $\alpha_{k,p}$ is the complex gain of the $p$th path,
$\a_{UT,k}(\theta_{k,p})$ is the UT antenna array response vector with the AoA $\theta_{k,p}$, and $\a_{BS}(\varphi_{k,p})$ is the BS antenna array response vector corresponding to AoD $\varphi_{k,p}$.
Specifically, under the uniform linear array (ULA) setup,
these vectors are given by
\begin{align}
    \a_{UT,k}(\theta_{k,p}) =  \frac{1}{\sqrt{N_k}} \left[
    1, e^{-j\frac{2\pi}{\lambda}d \sin(\theta_{k,p})},\ldots,\right.\nn\\
    \left. e^{-j(N_k-1)\frac{2\pi}{\lambda}d \sin(\theta_{k,p})}\right]^T\nn\\
    \a_{BS}(\varphi_{k,p})=\frac{1}{\sqrt{M}} \left[
    1, e^{-j\frac{2\pi}{\lambda}d \sin(\varphi_{k,p})},\ldots,\right.\nn\\
    \left. e^{-j(M-1)\frac{2\pi}{\lambda}d \sin(\varphi_{k,p})}\right]^T,
\end{align}
where $\lambda$ is the wavelength, and $d$ is the distance between the adjacent antennas.
For a narrow-band channel model, the channel response of the $k$th UT
can be expressed as
\begin{align}
	\H_{k}^{DL}= \sum_{p=1}^{N_P} \H_{k,p}^{DL}.
\end{align}
The channel covariance matrices at the BS and the UT sides are defined as
\begin{align}
    \R_{BS,k} &=\Eb \{ (\H_{k}^{DL})^H \H_{k}^{DL} \},\\
    \R_{UT,k} &= \Eb \{  \H_{k}^{DL} (\H_{k}^{DL})^H \}.
\end{align}

In this paper, we assume the uplink and downlink channels are reciprocal, i,e,
\begin{align}\label{eq:reciprocal}
	\H_{k}^{UL}= (\H_{k}^{DL})^T,
\end{align}
but the channel estimations are affected by noise.
The extension to channel non-reciprocity caused by the time difference and hardware imperfection is beyond the scope of this paper.

\subsection{Problem Statement} 
\label{sub:problem}
{An intuitive approach is extending existing pairwise PKG approaches via allocating orthogonal pilots among UTs.}
Firstly, in the downlink channel probing, the BS broadcasts the pilot signal, $\S^{DL}\in \Cb^{M\times M}$, and the received signal of the $k$th UT is given by
\begin{align}\label{eq:2}
    \Y_k^{DL} =  \H_k^{DL} \S^{DL}+ \N_k^{DL} ,
\end{align}
where $\N_k^{DL}\in\Cb^{N_k\times M}$ is the Gaussian noise at UT $k$.
To estimate the perfect CSI, the pilot signal should satisfy the orthogonality, i.e., $\S^{DL}(\S^{DL})^H = \I_{M}$.
By the least square (LS) estimation, UT $k$ estimates his downlink CSI as
\begin{align}\label{eq:3}
	\Z_k^{DL} = \Y_k^{DL} (\S^{DL})^H  =\H_k^{DL} + \N_k^{DL} (\S^{DL})^H.
\end{align}

Next, in the uplink channel probing, all the UTs send the pilot signals, $\S_k^{UL}\in \Cb^{N_k\times N},k \in \{1,2,\cdots, K\}$ to the BS simultaneously, where $N=\sum_k N_k$ is the total number of UTs' antennas.
The received signal of BS is
\begin{align}\label{eq:4}
	\Y^{UL} = \sum\limits_{k=1}^K \H_k^{UL} \S_k^{UL}+ \N^{UL}.
\end{align}
where $\N^{UL}\in\Cb^{N_k\times N}$ is the Gaussian noise at the BS.
The BS estimates the uplink CSI of UT $k$ as
\begin{align}\label{eq:5}
	\Z_k^{UL} = \Y^{UL}(\S_k^{UL})^H =\H_k^{UL} \S_k^{UL}(\S_k^{UL})^H + \nn\\
    \H_k^{DL} \sum\limits_{k'\ne k}^K {\S_{k'}^{UL}}(\S_k^{UL})^H + \N^{UL} (\S_k^{UL})^H.
\end{align}
To distinguish each UT, the pilot signals of different UTs are designed to satisfy the orthogonality requirement, i.e.,
\begin{align}
\S_i^{UL} (\S_j^{UL})^H = \left\{ {\begin{array}{*{20}{c}}
\I_{N_i}, &i = j\\
{\bf 0}_{N_i \times N_j}, &i \ne j.
\end{array}} \right.
\end{align}
Therefore, (\ref{eq:5}) can be further reduced to
 \begin{align}\label{eq:6}
	\Z_k^{UL} =\H_k^{UL}+ \N^{UL} (\S_k^{UL})^H.
\end{align}
When $\S^{DL}$ and $\S_k^{UL}$ are unitary, the noise $ \N_k^{DL} (\S^{DL})^H$ and $\N^{UL} (\S_k^{UL})^H$ have the same distribution as $\N_k^{DL}$ and $\N^{UL}$
{because Gaussian distribution is isotropic and thereby unitarily invariant.}

According to (\ref{eq:reciprocal}), the BS and UT $k$ obtain very similar channel estimations of $\Z_k^{UL} \approx \Z_k^{DL}$.
Then, they vectorize the estimations $\z_k^{UL} = \tvec(\Z_k^{UL})$ and
$\z_k^{DL} = \tvec(\Z_k^{DL})$, which are chosen as channel characteristics for key generation. By employing quantization, information reconciliation and privacy amplification, the BS and UT $k$ finally generate consistent secret key $\kappa_k$.

However, this intuitive approach has two issues as follows.
\begin{enumerate}
   \item Define the duration of one round of channel probing as $T_{p}= T_D+T_U +T_{Switch}$, where $T_{Switch}$ is the switching time from downlink to uplink, $T_D$ and $T_U$ are the pilot transmission time in the downlink and the uplink, respectively.
   This time needs to be deliberately kept smaller than the channel coherence time, so that BS and UTs can obtain highly correlated CSI in a TDD system. However, in this case, $T_{p} = (M+\sum_k N_k) \Delta T +T_{Switch}$, where $\Delta T$ is the symbol transmission time.
To distinguish different antennas of different UTs, the length of uplink pilots scales with the number of antennas $N_k$ as well as the number of UTs $K$.
   When $M$, $K$ and $N_k$ are large, it becomes very challenge to accomplish channel probing within the coherence time.

    \item {Because of the spatial correlation of the antennas, the elements of $\z_k^{UL}$ and $\z_k^{DL}$ are highly auto-correlated, resulting to long 0s and 1s in the quantized bit sequences. Traditionally, preprocessing approaches, e.g., PCA, are used to reduce the auto-correlation.
    However, due to the large scale of antennas at both BS and UTs in the future wireless communications, it is complicated to perform PCA algorithm for $\z_k^{UL}$ and $\z_k^{DL}$ with a large dimension of $M N_k$.
}

\end{enumerate}

To sum up, the core problems are how to reduce the length of pilots and the high dimension of channel matrix in the multi-user massive MIMO system.
Fortunately, literature and field measurements have shown that the beam domain channel matrix reveals the sparse property in typical scenarios \cite{Scaling_up,BDMA}.
Hence, we propose a new channel dimensionality reduction (CDR)-based key generation scheme to address the above problems.

\section{General CDR-Based Key Generation Scheme}

In massive MIMO channels, a few dominant elements contain the most relevant channel information. To reduce the dimensions, we first introduce the beam domain transform and then propose the corresponding key generation scheme. The achievable secret key rate in the proposed scheme is also derived.

\subsection{Beam Domain Transform} 
\label{sub:asymptotic_channel_properties}
Beam domain transform samples the original physical channel by two series of uniformly distributed beams/angles over $[0,2\pi]$, i.e., transmitting and receiving beams/angles.
According to \cite{8465975}, the beam domain channel is
\begin{align}\label{eq:6}
    \Ht_{k}^{DL} = \A_{UT,k}^H \H_{k}^{DL} \A_{BS},
\end{align}
where
\begin{align}
    \A_{UT,k} = \left[\a_{UT,k}(\theta_1),\a_{UT,k}(\theta_2),\ldots,\a_{UT,k}(\theta_{N_k})  \right]  \in \Cb^{N_k\times N_k}
\end{align}
and
\begin{align}
    \A_{BS} = \left[\a_{BS}(\varphi_1),\a_{BS}(\varphi_2),\ldots,\a_{BS}(\varphi_M)  \right]  \in \Cb^{M\times M}
\end{align}
are the sampling matrices at the $k$th UT and the BS, respectively. They satisfy that $ \A_{UT,k}^H \A_{UT,k} = \I$, $\A_{BS}^H \A_{BS}=\I $. The $(n,m)$-th element of $\Ht_{k}^{DL}$ represents the channel gains from AoD $\varphi_m$ to  AoA $\theta_n$, where $\varphi_m$ and $\theta_n$ are the $m$th and $n$th sample angles, which satisfy that $\sin(\varphi_m) = 2m/M-1$ and $\sin(\theta_n) = 2n/N_k-1$\cite{1033686}.
When the antenna spacing is half wavelength, i.e., $d = \lambda/2$,
the matrices $\A_{UT}$ and $\A_{BS}$ become the unitary discrete Fourier transform (DFT)
matrix, defined as \cite{7913686}
\begin{align}
    [\A_{UT,k}]_{n_1,n_2} = \frac{1}{\sqrt{N_k}} \exp(-j2\pi(n_1-1)(n_2-N_k/2)/N_k)\nn\\
    [\A_{BS}]_{m_1,m_2} = \frac{1}{\sqrt{M}} \exp(-j2\pi(m_1-1)(m_2-M/2)/M).
\end{align}

When the number of antennas tends to infinity,
the beam domain channel $\Ht_{k}^{DL}$ exhibits
{the spatial resolution as follows \cite{BDMA}.}
\begin{prop}\label{prop:1}
    When the number of antennas grows to infinity,
    the beam domain channel $\Ht_{k}^{DL}$ tends to $\G_{k}^{DL}$, i.e.,
    for arbitrary $n$ and $m$,
    \begin{align}
        \lim_{M,N_k\to \infty} [\Ht_{k}^{DL} - \G_{k}^{DL}]_{n,m} = 0,
    \end{align}
    where $\G_{k}^{DL} \in \Cb^{N_k\times M}$ is given by
    \begin{align}\label{eq:G}
        [\G_{k}^{DL}]_{n,m} = \sum_{p=1}^{N_P} \alpha_{k,p} \delta(\theta_{k,p} - \arcsin(2n/N_k-1))\nn\\
        \times\delta(\varphi_{k,p} - \arcsin(2m/M-1)).
    \end{align}
\end{prop}
\begin{IEEEproof}
    See Appendix \ref{sec:proof_of_proposition_prop:1}.
\end{IEEEproof}

\begin{rem}
From the definition of $\G_{k}^{DL}$ in \eqref{eq:G},
for each $n$ and $m$, there is at most one path $p$ simultaneously satisfying $\theta_{k,p} = \arcsin(2n/N_k-1)$ and $\varphi_{k,p} = \arcsin(2m/M-1)$.
This indicates that one element in $\G_{k}^{DL}$ represents channel gains from one AoD $\varphi_{k,p}$ to one AoA $\theta_{k,p}$ and
different elements represent channel gains corresponding to different AoAs and AoDs.
As there are $N_P$ paths, the number of non-zero entries in matrix $\G_k^{DL}$ is $N_P$. When the BS and UT $k$ are equipped with a large (but finite) number of antennas, the beam domain channel matrix $\Ht_{k,p}^{DL}$ can be approximated by $\G_{k}^{DL}$. In this case, $\Ht_k^{DL}$ is a very sparse matrix with $N_P$ dominant elements corresponding to the paths.
Moreover, these elements become independent with each other as long as these paths are independent.
\end{rem}

Define the beam domain channel covariance matrices at the BS and $k$th UT as
\begin{align}\label{eq:019}
    \Rt_{BS,k}  &= \Eb \{ (\Ht_{k}^{DL})^H \Ht_{k}^{DL} \}=\A_{BS}^H \R_{BS,k} \A_{BS},\nn\\
    \Rt_{UT,k}  &= \Eb \{  \Ht_{k}^{DL} (\Ht_{k}^{DL})^H \}=\A_{UT,k}^H \R_{UT,k} \A_{UT,k},
\end{align}
respectively.
When the number of antennas grows to infinity,
$\Rt_{BS,k}$ and $\Rt_{UT,k}$ tend to diagonal matrices with the diagonal elements given by
\begin{align}\label{eq:9}
    \lim_{M\to\infty} &[\Rt_{BS,k}]_{m,m}- \sum_{p=1}^{N_P} |\alpha_{k,p}|^2
        \delta(\varphi_{k,p} - \arcsin(2m/M-1)) \nn\\ &= 0,\nn\\
    \lim_{N_k\to\infty} &[\Rt_{UT,k}]_{n,n} -\sum_{p=1}^{N_P} |\alpha_{k,p}|^2 \delta(\theta_{k,p} - \arcsin(2n/N_k-1))\nn\\& = 0.
\end{align}
The $m$th diagonal element in $\Rt_{BS,k}$ represents the channel gains of the $m$th transmit beam ($\varphi_{k,p} = \arcsin(2m/M-1)$), and the $n$th diagonal element in $\Rt_{UT,k}$ represents the channel gains of the $n$th receive beam ($\theta_{k,p} = \arcsin(2n/N-1)$). The beam domain channel covariance matrices also reveal the angular resolution of the channel gains.

From the above analysis, one can observe that the representation in the beam domain channel brings the following benefits. Firstly, in the beam domain, the channel matrix reveals the sparse property, i.e., only a few elements contain the most channel information,
which reduce the dimension of channel estimation.
Secondly, as the number of antennas at the BS and UT increases, the elements of the channel matrix become mutually independent, reducing the redundancy, which is particularly desirable in secret key generation.
Thirdly, with a large number of antennas at the BS, the beam domain transform matrix at the BS $\A_{BS}$ is independent of UTs and we can use one identical matrix to transform channel matrices of different UTs into the beam domain. Such property is desirable in multi-user secret key generation.

\subsection{Key Generation Scheme Based on CDR} 
\label{sec:CDR-KGS}
\begin{figure}[!t]
\centering
{\includegraphics[width=0.35\textwidth]{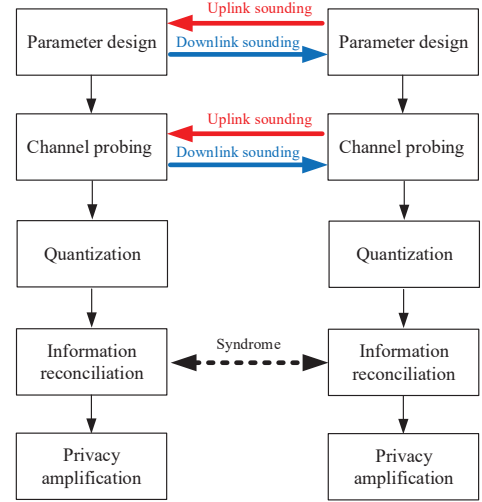}}
\caption{Flow diagram of CDR based secret key generation scheme.}
\label{fig:scheme}
\end{figure}

In the beam domain, only a few elements contain the most channel information,
which motivates us to propose a general framework for multi-user secret key generation, as portrayed in Fig.~\ref{fig:scheme}.
The UT and the BS transmit sounding signals to acquire the statistical CSI.
During the parameter design stage, the BS and UTs design precoding and receiving matrices based on the statistical CSI, in order to reduce the dimension of channel estimation.
Then, the BS and UTs estimate the effective channel parameters with high correlations. After quantization, the information reconciliation and privacy amplification procedures are used to generate consistent and private secret keys, similarly with the point-to-point secret key generation. Information reconciliation and privacy amplification are thus not particularly designed or optimized in this paper.

In this paper, we focus on the first two steps, i.e., parameter design and channel probing, which is relatively different from that in the point-to-point secret key generation.
\begin{enumerate}
  \item {\textit{Parameter design:}} In this step, BS and UTs design the precoding and receiving matrix according to their statistical CSI.
Firstly, in the uplink, each UT employs one antenna to transmit the sounding signals.
{Then, the BS estimates the covariance matrix and designs the precoding matrix $\P_k$
with equal power allocation $\P_k^H\P_k = \I$.}
Next, in the downlink, the BS employs the precoding matrix to transmit the downlink sounding signals and
{UTs estimate the statistical CSI information of the covariance matrix at the UT side and design the receiving matrix $\C_k$ satisfying $\C_k^H\C_k = \I$.}
    \item {\textit{Channel probing:}}In this step, BS and UTs probe the channel alternatively and construct the reciprocal channel characteristics with the help of the precoding and receiving matrix.
 Firstly, the BS transmits the downlink pilot signals by the precoding matrix $\P$ and UTs preprocess the received signals by the  matrix $\C^H$ to obtain the reciprocal channel parameters.
 Next, each UT employs the matrix $\C^*$ to transmit the pilot signals. The BS utilizes the precoding matrix $\P$ to preprocess the received signals and estimate the effective channel.
\item{
{Quantization: After channel probing, using channel quantization alternating (CQA) scheme~\cite{Wallace2009Key},
the BS and each UT quantize the effective channel measurement to generate
the initial secret keys.}
}
\end{enumerate}

{
\begin{rem}
	The parameter design is proposed specifically for multi-user
	massive MIMO secret key generation, which is used to reduce the channel
	estimation dimension, as well as the inter-user interference.
	The dimensions of $\P_k$ and $\C_k^H$ are $M\times M_e$ and $N_e\times N_k$ respectively, where $M_e$ and $N_e$ are the reduced dimensions at the BS and UT,
	which are approximately equal to the number of paths $N_P$, far smaller than $M$ and $N_k$.
	Then, we only need to estimate the effective channel with $N_e\times M_e$,
	significantly reducing the channel estimation dimension.
	Moreover, from the analysis of channel characteristics,
	one can observe that the channel gains of one UT are concentrated within a few beams (directions),
               which has the potential to separate different UTs by different beams.
	In addition, the precoding and receiving matrices are determined by
	the statistical CSI, which can be obtained by some time and frequency resources \cite{BDMA}.

Once the parameter design is completed, the BS and UTs can perform multiple
channel probing rounds. Each channel probing round, including the uplink
and downlink channel sounding, should be completed within one coherence time slot,
where the instantaneous CSI keeps constant~\cite{ZHANG2020Frontier}.
Different channel probing rounds operate in different channel coherence time slots,
and the instantaneous CSI varies along time, resulting in
the variation of the generated secret keys.
\end{rem}
}

\begin{figure}[!t]
\centering
\subfigure[Downlink]{\includegraphics[width=0.4\textwidth]{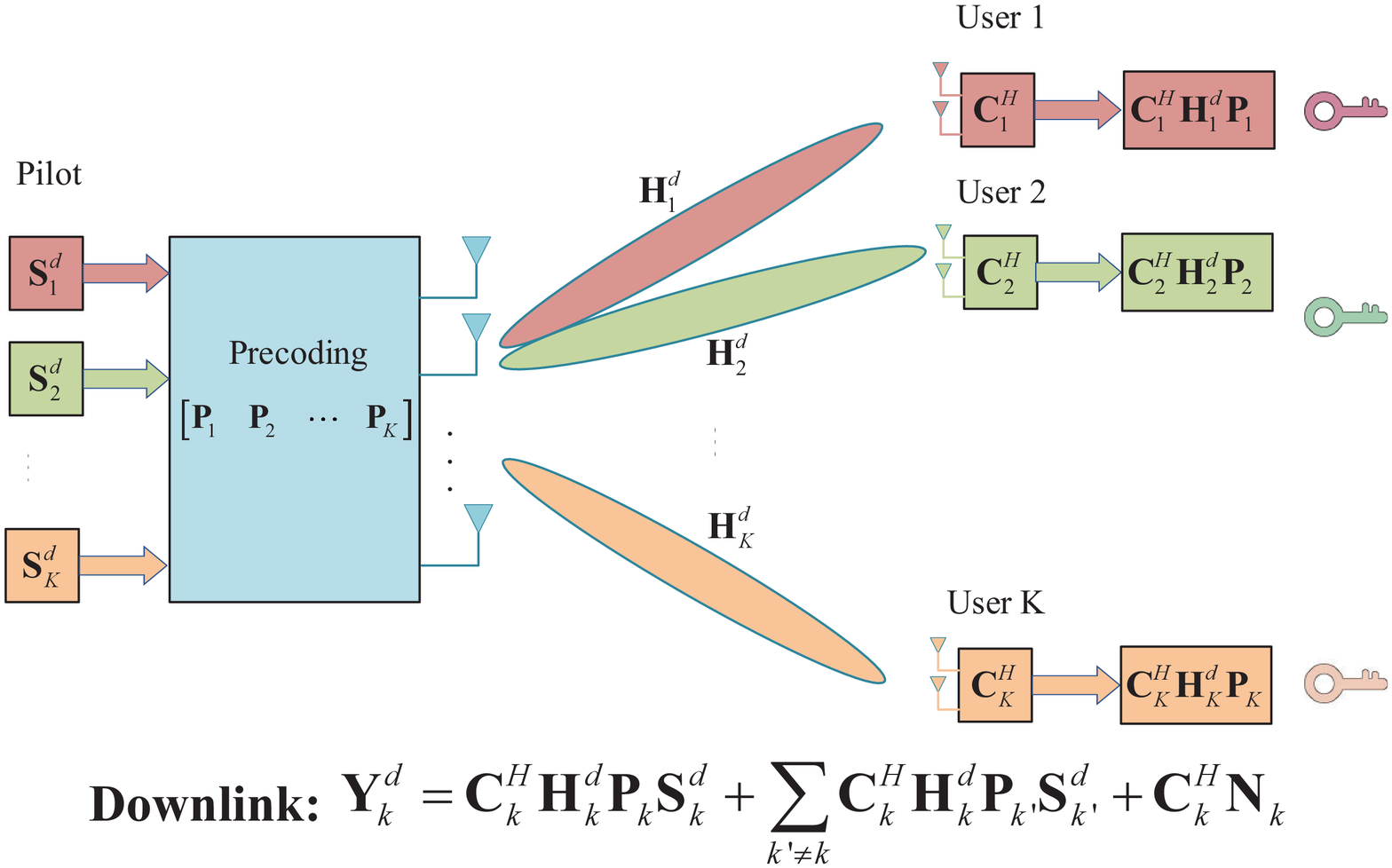}\label{fig:1}}
\subfigure[Uplink]{\includegraphics[width=0.4\textwidth]{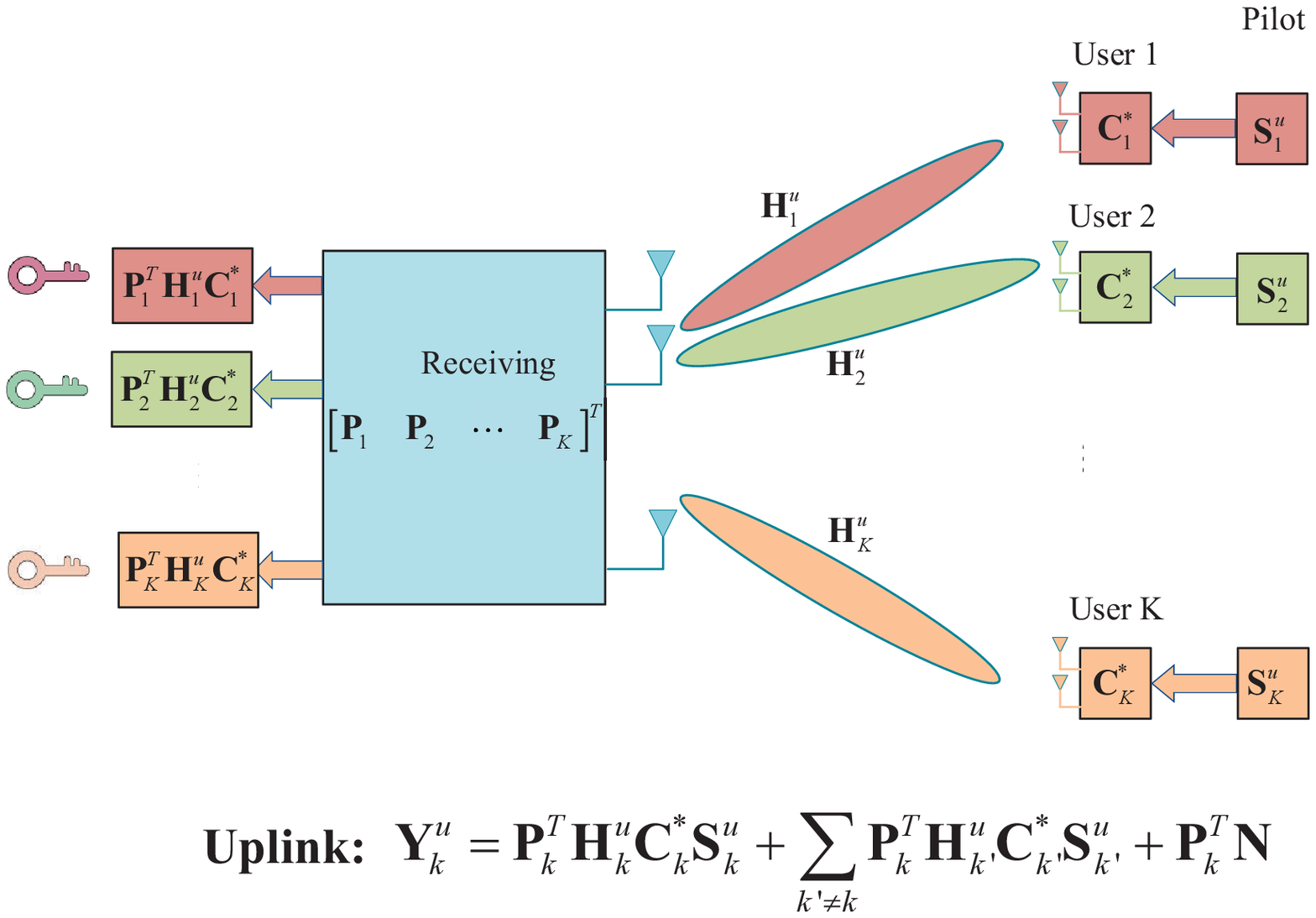}\label{fig:2}}
\caption{Channel probing in secret key generation scheme.}\label{fig:mmW}
\end{figure}

The downlink and uplink probing process is illustrated in Fig.~\ref{fig:mmW}.
Let $\P= [\P_1,\P_2,\cdots,\P_K]$ and $\C^H= [\C_1^H,\C_2^H,\cdots,\C_K^H]$ denote the precoding and receiving matrices in the downlink transmission, respectively.
Then, $\C^*$ and $\P^T$ are used as precoding and receiving matrices in the uplink transmission.

Specifically, let $\S_k^{DL}\in \Cb^{M_e\times T_D}$ denote the downlink pilot from BS to UT $k$ within $T_D$ time slots, which satisfies $\S_{k}^{DL}(\S_k^{DL})^H = \I$.
After multiplying the pilot signals by the precoding matrix $\P_k$, the BS transmits the summation of all the UTs. Then, UT $k$ multiplies the received signal by the receiving matrix $\C_k^H$, given by
\begin{align}\label{eq:20}
	\Y_k^{DL} = \C_k^H \H_k^{DL} \P_{k} \S_{k}^{DL}+ \C_k^H \H_k^{DL} \sum_{k'\ne k} \P_{k'} \S_{k'}^{DL} + \C_k^H \N_k.
\end{align}
By the LS estimation, UT $k$ estimates the downlink CSI as
\begin{multline}\label{eq:21}
	\Z_k^{DL} = \Y_k^{DL} (\S_k^{DL})^H = \C_k^H \H_k^{DL} \P_{k} \\
 +\C_k^H \H_k^{DL} \sum_{k'\ne k}\P_{k'}\S_{k'}^{DL} (\S_k^{DL})^H + \C_k^H\N_k (\S_k^{DL})^H.
\end{multline}

In the uplink transmission, let $\S_k^{UL}\in\Cb^{N_e\times T_U}$ denote the pilot transmitted by UT $k$ within $T_U$ time slots, satisfying $\S_{k}^{UL} (\S^{UL}_k)^H = \I$.
The $k$th UT transmits pilot signals by the matrix $\C_k^*$, and the BS receives the summation of all the UTs' signals.
Then, multiplying by the receiving matrix $\P_k^T$, which is transpose of the precoding matrix in the downlink, the received signal of UT $k$ at the BS can be expressed as
\begin{align}\label{eq:18}
	\Y^{UL}_k = \P_k^T \H_{k}^{UL}\C_{k}^*\S_{k}^{UL}+\P_k^T \sum_{k'\ne k}\H_{k'}^{UL}\C_{k'}^*\S_{k'}^{UL} + \P_k^T \N
\end{align}
where $\Y^{UL}_k\in\Cb^{M_e\times T_U}$ is the received signals of time length $T_U$.
By employing the LS estimation, the estimated effective channel of UT $k$ can be expressed as
\begin{multline}\label{eq:19}
	\Z_k^{UL} = \Y^{UL} (\S^{UL}_k)^H = \P_k^T\H_{k}^{UL}\C_{k}^* \\
    +\P_k^T \sum_{k'\ne k} \H_{k'}^{UL}\C_{k'}^*\S_{k'}^{UL} (\S^{UL}_k)^H + \P_k^T\N (\S^{UL}_k)^H.
\end{multline}
\begin{rem}
In the uplink and downlink transmission, the BS and UTs vectorize the estimated effective channel matrices $\Z_k^{UL}$ and $\Z_k^{DL}$ and employ them to generate the secret key, where the reciprocal component between the BS and UT $k$ is $\C_k^H \H_k^{DL} \P_k$ with a small dimension of $N_e \times M_e$.
In this way, the dimension of channel characteristics is reduced by $\eta = \frac{M\times N_k}{N_e \times M_e}$ times. The dimensions of $M_e$ and $N_e$ are very small compared with the number of antennas, therefore
the reduction $\eta$ is very significant. In addition, the duration of one round of channel probing $T_{p}$ is reduced from $(M+\sum_k N_k) \Delta T +T_{Switch}$ to $(M_e+N_e) \Delta T +T_{Switch}$.
\end{rem}

Let $\S_{k'k}^{DL} = \S_{k'}^{DL} (\S_{k}^{DL})^H$ (or $\S_{k'k}^{UL} = (\S_{k'}^{UL} (\S^{UL}_k)^H)^T$) represent the covariance matrix of the downlink (or uplink) pilot signals, where $\S_{kk}^{DL} = \S_{kk}^{UL} = \I$. Moreover, as the uplink and downlink channel are reciprocal, for the simplicity of notation, the downlink channel $\H_k^{DL}$ is denoted as $\H_k$, and the uplink channel is $\H_k^{UL}=(\H_k)^T$. Then, the vectorized channel matrices can be expressed as
\begin{align}
    \z_k^{DL} =& \tvec(\Z_k^{DL}) = \sum_{k'} \left(\left(\P_{k'}\S^{DL}_{k'k}\right)^T \otimes \C_k^H\right)\tvec(\H_k) \nn\\&+ \tvec(\C_k^H\N_k ) \\
    \z_k^{UL} =& \tvec((\Z_k^{UL})^T) = \sum_{k'}\left(\P_k^T \otimes\S_{k'k}^{UL}\C_{k'} \right) \tvec(\H_{k'}) \nn\\ &+ \tvec(\N^T \P_k).
\end{align}
Although the uplink and downlink channel of the $k$th UT are identical, the interference terms between UTs are not reciprocal.
Specifically, in the downlink transmission, the interference received at the $k$th UT is the summation of the transmitted signals of the $k'$th UT propagating through the channel of the $k$th UT $\H_{k}$, while in the uplink transmission, the interference received at the BS for the $k$th UT is the summation of the transmitted signals of the $k'$th UT propagating through the channel of different UTs $\H_{k'}$.
Thus, the interference will also reduce the agreement.

\subsection{Secret Key Rate}
\label{sub:secret_key_rate}
When the BS communicates with one UT, other UTs are potential non-colluding curious users\footnote{{This paper focuses on the non-colluding scenario, where no information is shared among the curious users.
The colluding case, where curious users share their received signals with each other, can be studied in the future.
}}.
Under the TDD operation, each UT cannot transmit and receive signals at the same time. The $i$th UT only has the channel observation in the downlink transmission.
Thus, the key rate is the minimum mutual information given other UT's observations. The number of secure bits for the link from the BS to UT $k$ in the mutual information can be expressed as \cite{Liang2009}
\begin{align}\label{eq:27}
    I_{k} = \min_{i\neq k} I (\z_k^{DL};\z_k^{UL} | \z_i^{DL}  ).
\end{align}

\begin{rem}
We assume that the distance between the BS and each UT is several orders of magnitude larger than the wavelength, i.e., there is no UT close to the BS.
Then, in the uplink transmission, the channel from one UT to the BS is independent of that from one UT to another UT. Moreover, as UTs transmit signals at the same time and frequency block, they cannot receive signals from other UTs. Thus, the secret key rate is the minimum mutual information between $\z_k^{DL}$ and $\z_k^{UL}$ on the condition of $\z_i^{DL}$.
\end{rem}

When the channel estimations of different UTs are independent, the secret key rate degrades to
\begin{align}
    I_k = I (\z_k^{DL};\z_k^{UL} ).\label{eq:degrade}
\end{align}
In massive MIMO communications, when the beam domain channels of different UTs are non-overlapping, i.e., the channel covariance matrices at the BS are orthogonal, given by
\begin{align}
     \Rt_{BS,k}\Rt_{BS,i} = {\bf 0},\quad k\neq i,
 \end{align}
the channel vectors of UT $k$ and UT $i$ are independent. Then, the secret rate $I(\z_k^{DL};\z_k^{UL}|\z_i^{DL})$ can be degraded as $I(\z_k^{DL};\z_k^{UL})$ and no secret keys are leaked to potential curious UTs~\cite{6584929}.

When the beam domain channels are overlapping, we should consider the information leakage to other UTs. But we can always select non-overlapping beams for different UTs and then the selected channel information is independent. Thus, we can also use (\ref{eq:degrade}) to calculate the secret key rate. The overlapping case will be discussed in more detail in Section~\ref{sub:security_analysis}.


Denote the precoding and receiving matrices in the beam domain as $\Pt_k = \A_{BS}^H\P_{k}$ and $\Ct_k = \A_{UT,k}^H\C_k$, respectively.
Let $\V_k = \Lambdam_k^{1/2}\Big(\sum_{k'}(\Pt_{k'}\S^{DL}_{k'k})^T \otimes \Ct_k^H\Big)^H$ and $\V_{kk'} = \Lambdam_{k'}^{1/2} \Big(\Pt^T_k\otimes \S^{UL}_{k'k}\Ct^H_{k'}\Big)^H$,
where $\Lambdam_k = \Eb\{\tvec(\Ht_k)\tvec(\Ht_k)^H\}$ is the full correlation of the beam domain channel.
We can compute the secret key rate of UT $k$ as follows.
\begin{thm}\label{thm:1}
    When the channels of different UTs become independent, the secret key rate of the $k$th UT is given by
    \begin{align}\label{eq:53}
        &I(\z_k^{DL};\z_k^{UL}) \nn\\&= \!- \log\! \det\! \Bigg(\!\I - \V_{kk}\!  \left(\! \sum_{k'} \V_{kk'}^H \V_{kk'}  \!+\! \Big(\P^T_k\P^*_k \otimes \I_{T_U}\!\Big)\!\!\right)^{-1}\!\!\!\! \V_{kk}^H\nn\\
    &\quad \times  \V_k \left(  \V_k^H\V_k + \I_{T_D} \otimes \C_k^H \C_k \right)^{-1}\V_k^H \Bigg).
    \end{align}
\end{thm}
\begin{IEEEproof}
    See Appendix \ref{sec:proof_of_theorem_thm:1}.
\end{IEEEproof}
\begin{rem}
    Note that when the number of UTs is one (i.e., there is only one UT), the secret key rate reduced to the single user secret key rate, which is a special case of \eqref{eq:53}. Moreover, the secret key rate \eqref{eq:53} is complicated, which depends on the precoding and receiving vectors, as well as the covariance matrices of pilot signals.
    To estimate the effective CSI, the pilot signals of one UT should be orthogonal, i.e., $\S_{kk}^{DL} = \I$ and $\S_{kk}^{UL} = \I$ \cite{How}.
    When the pilot signals between UTs are orthogonal, i.e., $\S_{k'k}^{DL} = {\bf 0}$ and $\S_{k'k}^{UL} = {\bf 0}$, there is no interference between UTs. The secret key rate can be simplified as
    \begin{align}\label{eq:36}
        &I(\z_k^{DL};\z_k^{UL}) =  -\log \det  \Bigg(  \I - \Lambdam_k^{1/2} \Big(\Pt^* _k\otimes \Ct_{k}\Big)  \nn\\
        &\times \Bigg( \I + \Big(\Pt^T_k\otimes \Ct^H_{k}\Big)
     \Lambdam_{k} \Big(\Pt^*_k \otimes \Ct_{k}\Big) \Bigg)^{-1}\Big(\Pt^T_k\otimes \Ct^H_{k}\Big)
    \Lambdam_k \nn\\
    &  \times
     \Big(\Pt_{k}^* \otimes \Ct_k\Big) \Bigg(\I +\Big(\Pt_{k}^T \otimes \Ct_k^H\Big) \Lambdam_k\Big(\Pt_{k}^* \otimes \Ct_k\Big)\Bigg)^{-1}\nn\\
    &\times \Big(\Pt_{k}^T \otimes \Ct_k^H\Big) \Lambdam_k^{1/2}\Bigg).
    \end{align}
    However, for the orthogonal pilots between UTs, the pilot overhead scales with the number of UTs, which is quite large in multi-user communication systems.
    In general, due to the short coherent time, employing the orthogonal pilots between users is impractical.
    Alternatively, pilot signals can be reused between UTs, i.e., $\S_{k'k}^{DL} = \I$ and $\S_{k'k}^{UL} = \I$. Under this condition, there exists inter-user interference and we will design the precoding and receiving matrices to reduce it.
\end{rem}

\section{Optimization Design with Pilot Reuse}

In this section, we consider the CDR-based secret key generation scheme design under the pilot reuse case, where different UTs transmit the identical pilot signals.
In this case, we first design the precoding and receiving matrices maximizing the secret key rate and then analyze the security when the channels of different UTs are correlated.




\subsection{Design of Precoding and Receiving Matrices} 
\label{sec:design_maximizing_the_sum_secret_key_rate}

Under the pilot reuse case, the inter-user interference will affect secret key agreement. Therefore, we need to design the precoding and receiving matrices  to maximize the sum secret key rate.
Generally, the precoding and receiving matrices contain the transmit directions as well as the transmitted power on each direction.
To reduce the interference, we focus on the transmit direction design and consider the equal power allocation of each direction,
which can be expressed as
\begin{align}\label{eq:p0}
	\max_{\Pt_k,\Ct_k} \quad & R_{\text{sum}} = \sum_k I(\z_k^{DL};\z_k^{UL}) \nn\\
	\st \quad &  \Pt_k^H \Pt_k = \I \nn\\
	& \Ct_k^H \Ct_k = \I,
\end{align}
where the secret key rate is calculated as
    \begin{align}\label{eq:37}
        &I(\z_k^{DL};\z_k^{UL}) =  -\log \det \left(  \I - \Lambdam_k^{1/2} \Big(\Pt^T_k\otimes \Ct^H_{k}\Big) \right.  ^H\nn\\
        &   \times \left(\I + \sum_{k'}\Big(\Pt^T_k\otimes \Ct^H_{k'}\Big)
     \Lambdam_{k'} \Big(\Pt^T_k \otimes \Ct^H_{k'}\Big) ^H\right)^{-1} \nn\\
    &  \times \Big(\Pt^T_k\otimes \Ct^H_{k}\Big)
    \Lambdam_k\Big(\sum_{k'}(\Pt_{k'})^* \otimes \Ct_k\Big)\nn\\
    &\times  \left( \I + \Big(\sum_{k'}(\Pt_{k'})^T \otimes \Ct_k^H\Big) \Lambdam_k\Big(\sum_{k'}(\Pt_{k'})^* \otimes \Ct_k\Big)  \right)^{-1}\nn\\
    &\left.\times\Big(\sum_{k'}(\Pt_{k'})^T \otimes \Ct_k^H\Big) \Lambdam_k^{1/2}\right).
    \end{align}
{Although we consider the sum secret key rate maximization,
the key rate differences among UTs are not large.
As we consider the equal power allocation for different beams,
most UTs can achieve similar secret key rates.}

{
As in the objective function \eqref{eq:37}, the optimization matrices
$\Pt_k$ and $\Ct_k$ are involved both inside and outside the matrix inversion operation, the function \eqref{eq:37} is not convex on $\Pt_k$ and $\Ct_k$,
resulting in the non-convex problem \eqref{eq:p0},
}
which is difficult to solve globally.
In order to reduce the computational complexity and lower the pilot overhead, we utilize the interference neutralization approach~\cite{6584929} to mitigate the interference, i.e., for arbitrary matrix $\Ct_{k'}$ ($k'\neq k$), the precoding matrix $\Pt_k$ satisfies
\begin{align}\label{eq:48}
    (\Pt_k^T\otimes \Ct_{k'}^H) \Lambdam_{k'} = {\bf 0},\quad k'\neq k.
\end{align}
This constraint indicates that the precoding matrix $\Pt_k$ can eliminate the inter-user interference.
Note that when the channel beams of different users are non-overlapping,
the precoding matrices corresponds to different beams, and thus we have
\begin{align}
    \Pt_k^H \Rt_{BS,k'}= {\bf 0},\quad k'\neq k.
\end{align}
Therefore, the constraint \eqref{eq:48} can be easily satisfied.

Under this constraint, problem \eqref{eq:p0} maximizing the sum secret key rate can be decomposed into the following subproblems maximizing the secret key rate of each user
\begin{align}\label{eq:prob1}
    \min_{\Pt_k,\Ct_k} \  & \log \det \Big( \I - \Lambdam_k^{1/2} \Big(\Pt^* _k\otimes \Ct_{k}\Big)  \Big(\I + \Big(\Pt^T_k\otimes \Ct^H_{k}\Big)
     \Lambdam_{k} \nn\\
     &\times\Big(\Pt^*_k \otimes \Ct_{k}\Big) \Big)^{-1}\Big(\Pt^T_k\otimes \Ct^H_{k}\Big) \Lambdam_k\Big(\Pt_{k}^* \otimes \Ct_k\Big) \nn\\ &\hspace{-0.9cm} \times \Big(\I +\Big(\Pt_{k}^T \otimes \Ct_k^H\Big) \Lambdam_k\Big(\Pt_{k}^* \otimes \Ct_k\Big)\Big)^{-1}\Big(\Pt_{k}^T \otimes \Ct_k^H\Big) \Lambdam_k^{1/2}\Big) \nn\\
    \st \quad & \Pt_k^H \Pt_k = \I \nn\\
     &\Ct_k^H \Ct_k = \I .
\end{align}
\begin{rem}
The secret key rate in \eqref{eq:prob1} is equal to that  using orthogonal pilots in \eqref{eq:36}. This means that when the precoding matrix $\Pt_k$ satisfies condition \eqref{eq:48}, UTs reusing the identical pilot signals approaches the performance with orthogonal pilot signals.
    Both schemes can mitigate the inter-user interference. The difference is that orthogonal pilot scheme uses orthogonal pilot signals to separate different UTs, which requires large pilot overhead, while interference neutralization scheme designs the precoding matrices to eliminate the interference, which is independent of pilot signals between UTs.
\end{rem}

Note that in problem \eqref{eq:prob1}, the secret key rate depends only on the Kronecker product $\Pt^*_k\otimes\Ct_{k}$.
Define $\U_k = \Big(\Pt^*_k\otimes\Ct_{k}\Big)$, which is also a tall unitary matrix.
We first consider the matrix $\U_k$ design maximizing the secret key rate, and then, we construct the precoding and receiving matrices satisfying the interference neutralization constraint.
The matrix $\U_k$ design problem can be expressed as
\begin{align}\label{eq:prob2}
    \min_{\U_k} \ & \log\det\left( \I - \left( \Lambdam_k^{1/2} \U_k  \left(\I + \U_k^H \Lambdam_{k} \U_k \right)^{-1}\U_k^H  \Lambdam_k^{1/2}\right)^2 \right)
     \nn\\
    \mbox{s.t.} \ & \U_k^H \U_k = \I .
\end{align}
The solution of problem \eqref{eq:prob2} is obtained as follows:
\begin{thm}\label{thm:2}
    The optimal $\U_k$ maximizing the secret key rate is
    \begin{align}\label{eq:U}
        \U_k = \begin{bmatrix}
            \e_{\eta_1} & \e_{\eta_2} & \cdots & \e_{\eta_{M_eN_e}}
        \end{bmatrix},
    \end{align}
    where $\e_i = [0,0,\cdots,0,1,0,\cdots,0]$ is a unit vector with the $i$th unit element and $\eta_i$ is the index of the $i$th sorted eigenvalue of matrix $\Lambdam_k(\I+\Lambdam_k)^{-1}$.
    The optimal rate is
    \begin{align}\label{eq:44}
        R_k = -\sum_{i=1}^{M_e N_e }\log \lambda_i \Big(\I - \Lambdam_k^{2}  (\I +  \Lambdam_k)^{-2}  \Big).
    \end{align}
\end{thm}
\begin{IEEEproof}
    See Appendix \ref{sec:proof_of_proposition_thm:1}.
\end{IEEEproof}
\begin{rem}
    To maximize the secret key rate of the $k$th UT, the optimal $\U_k$ consists of the unit vectors corresponding to the sorted diagonal elements in $\Lambdam_k$.
    However, as $\U_k$ has the structure $\Pt^*_k\otimes\Ct_{k}$, in general cases, it cannot satisfy the optimal condition \eqref{eq:U}.
    Next, we will employ the channel properties and Theorem \ref{thm:2} to construct the precoding and receiving vectors $\Pt_k$ and $\Ct_k$.
\end{rem}

As $\U_k$ is consist of vectors $\e_i$, the beam domain precoding and receiving matrices $\Pt_k$ and $\Ct_k$ have the similar structure, which only need to select corresponding beams.
This indicates that the optimal precoding and receiving matrices $\P_k$ and $\C_k$ are consist of the eigenvectors of the channel covariance matrices, i.e., the precoding matrix $\P_k$ is a sub-matrix of $\A_{BS}$, while the receiving matrix $\C_k$ is a sub-matrix of $\A_{UT,k}$.

Moreover, recalling Proposition 1, as the number of antennas tends to infinity, the beam domain channel matrix $\Ht_k$ approaches the matrix $\G_k$, where different elements represent the channel gains from different AoDs to different AoAs. The channel gains are concentrated in a few elements in $\G_k$.
Specifically, suppose that there are $N_P$ paths, corresponding to $N_P$ AoAs and $N_P$ AoDs. Then, the BS selects the strongest $N_P$ beams, i.e., the precoding matrix $\Pt_k$ is given by
\begin{align}\label{eq:pt}
    \Pt_k = \begin{bmatrix}
        \e_{\eta_{t,k,1}} & \e_{\eta_{t,k,2}} & \cdots & \e_{\eta_{t,k,N_P}}
    \end{bmatrix}
\end{align}
where $\eta_{t,k,1}$ is the index of the sorted eigenvalue of matrix $\R_{BS,k}$.
Similarly, UT $k$ selects the strongest $N_P$ receiving directions, i.e., the receiving matrix $\Ct_k$ is given by
\begin{align}\label{eq:ct}
    \Ct_k = \begin{bmatrix}
        \e_{\eta_{r,k,1}} & \e_{\eta_{r,k,2}} & \cdots & \e_{\eta_{r,k,N_P}}
    \end{bmatrix}
\end{align}
where $\eta_{r,k,1}$ is the index of the sorted eigenvalue of matrix $\R_{UT,k}$.
The number of paths $N_P$ is relatively small, and $M_e$ and $N_e$ can be chosen equal to the number of paths.
Using the precoding and receiving matrices, we can construct $\U_k = \Pt_k^* \otimes \Ct_k$ to obtain the $N_P^2$ elements in $\Lambdam_k$, which contains the channel information of the $N_P$ paths.
This approach can significantly reduce the pilot overhead and fits well to  massive MIMO channel model and precoding \cite{6717211}.

\subsection{Security Analysis with Overlapping Beams}
\label{sub:security_analysis}

In the above analysis, we assume that the channel matrices of different UTs are independent. This assumption can be easily satisfied for the non-overlapping case where the channel beams of different UTs are non-overlapping.
However, different UTs may have overlapping beams in reality. For example, when two UTs are close to each other, part of their channels may suffer the same propagation paths, resulting in the overlapping beams, i.e.,
\begin{align}
    \Rt_{BS,k}\Rt_{BS,k'} \neq {\bf 0},\quad k'\neq k.
\end{align}
Since the channels of the overlapping beams between $\Rt_{BS,k}$ and $\Rt_{BS,k'}$ are highly correlated, the independent assumption does not hold any more. Thus, the information leakage should be considered for the overlapping case design.

Next, we analyze the information leakage ratio for the overlapping case.
Note that for the $k$th UT, the beam domain channel elements in $\Ht_k$ are statistically independent and thus the secret key rate can be expressed as the summation of the key rate of each beam. Moreover, as UTs are assumed as non-colluding curious users, the information leakage is determined by the UT with the highest correlation.
Thus, we focus on the information leakage on one overlapping beam with two UTs as an example.

Suppose that both UT $1$ and UT $2$ occupy the identical beam $b$ at the BS. Denote the channel gains from beam $b$ at the BS to the dominant beam at UT $1$ (or UT $2$) as $h_1$ (or $h_2$). Assume that both $h_1$ and $h_2$ have the unit attenuation power, i.e., $\Eb\{h_1^2\}=\Eb\{h_2^2\} = 1$.
Define the information leakage ratio as $\gamma = (R_h-R_1)/R_h$, where $R_h$ is the secret key rate with the independent channel assumption and $R_1$ is the key rate when the channels are correlated.
Since the correlation reduces the secret key rate, we always have $R_h \ge R_1$.
Then, we can calculate $\gamma$ as follows.
\begin{thm}\label{thm:3}
	The information leakage ratio can be expressed as
	\begin{align}\label{eq:gamma}
			\gamma = 1- \frac{\log\frac{((1+\sigma^2)^2-\rho^2)^2}
		    {(1+\sigma^2)(\sigma^6+3 \sigma^4-2 \sigma^2\rho^2 + 2 \sigma^2)}}
			{\log\frac{(1 + \sigma^2)^2}{\sigma^2 (2 + \sigma^2)}},
	\end{align}
	where $\sigma^2$ is the noise variance and  $\rho$ is the cross channel correlation defined as
\begin{align}
\rho=\frac{\Eb\{h_1 h_2\}}{\Eb\{h_1^2\}\Eb\{h_2^2\}}={\Eb\{h_1 h_2\}}.
\end{align}

\end{thm}
\begin{IEEEproof}
	See Appendix \ref{sec:proof_of_proposition_thm:3}.
\end{IEEEproof}

\begin{rem}
	The information leakage ratio given by \eqref{eq:gamma} is complicated, depending on the correlation $\rho$ as well as the noise variance $\sigma^2$.
	Next, we will consider a special case. From Appendix \ref{sec:proof_of_proposition_thm:3}, when $\rho$ is $1$ or $-1$, the information leakage is the highest, which can be calculated as
	\begin{align}
		\gamma = \frac{\log\frac{3 + 2 \sigma^2}
    {\sigma^2(6 + 11 \sigma^2 + 6 \sigma^4 + \sigma^6)}}
		{\log\frac{(1 + \sigma^2)^2}{\sigma^2 (2 + \sigma^2)}}.
	\end{align}
	For high SNRs (low $\sigma^2$), as $\sigma^2$ tends to $0$,
	the information leakage ratio becomes
	\begin{align}
		\lim_{\sigma^2\to 0} \gamma = 1,
	\end{align}
	which indicates that the secret key rate goes to zero and vanishes.
\end{rem}

From the above analysis, one can observe that the channel correlation is mainly caused by the overlapping channel beams. Further, when the channel of the overlapping beams is highly correlated, the information leakage ratio tends to $1$. This result reveals that the correlated overlapping beams provide little secret key rate. Therefore, in the multi-user secret key generation, when two UTs have overlapping channel beams,
the BS should allocate \textit{non-overlapping transmitting beams} to different UTs,
i.e., the precoding vector $\Pt_k$ satisfies
\begin{align}\label{eq:47}
   \Pt_k \Rt_{BS,k'} = {\bf 0} , \quad k'\neq k.
\end{align}
This indicates that the allocated transmitting beams for the $k$th UT are not overlapping with the channel beams of other UTs.
Under this condition, the constraint \eqref{eq:48} is satisfied, and the channel matrices of the allocated beams for different UTs are independent.

\subsection{A Holistic Parameter Design Algorithm}
Combined with the result of above security analysis, we propose a holistic parameter design algorithm as illustrated in Algorithm \ref{alg:1}.

\begin{algorithm}[h]
\caption{Parameter design.}
\label{alg:1}
\begin{algorithmic}[1]
\Require $\R_{BS,k}$ and $\R_{UT,k}$
\Ensure $\P_{k}$ and $\C_k$
\State \textbf{At the BS side:}
\For{$k=1:K$}
\State  Calculate the beam domain channel covariance matrix $\Rt_{BS,k}$ according to \eqref{eq:019}.
\State Select the strongest non-overlapping beams $\Pt_k$ according to \eqref{eq:pt} and \eqref{eq:47}.
\State Construct the precoding matrix $\P_k=\A_{BS}\Pt_k$.
\EndFor
\State \textbf{At the UT side:}
\For{$k=1:K$}
\State  Calculate the beam domain channel covariance matrix $\Rt_{UT,k}$ according to \eqref{eq:019}.
\State Select the strongest beams $\Ct_k$ according to \eqref{eq:ct}.
\State Construct the receiving matrix $\C_k=\A_{UT}\Ct_k$.
\EndFor
\end{algorithmic}
\end{algorithm}

With the help of designed parameters, the BS and UTs can extract reciprocal channel information of the non-overlapping beams, on which the channels of different UTs are independent.
Therefore, we can complete the following key generation steps using the same approach as described in Section\ref{sec:CDR-KGS}.

It is noteworthy that the design of the matrices $\Pt_k$ and $\Ct_k$ depends on the statistical CSI $\Rt_{BS,k}$ and $\Rt_{UT,k}$. As the statistical CSI changes on a larger time scale than the instantaneous CSI, it is not necessary to design $\Pt_k$ and $\Ct_k$ for each secret key generation round. After designing $\Pt_k$ and $\Ct_k$, they can be used to generate secret keys until the statistical CSI changes.
Also note that an offline design is possible. Depending on the statistical CSI scenario, we can choose the corresponding pilots.

\section{Numerical Results} 
\label{sec:numerical_results}
In this section, we employ the numerical results to illustrate the performance of secret key generation in multi-user massive MIMO wireless communication systems.
A BS, equipped with $M=128$ antennas, simultaneously communicates with $K=6$ UTs, each with $N_k=4$ antennas. Here, we focus on massive antennas at the BS, which significantly affect the performance for the multi-user case. We consider the physical channel model, where there are $N_P=6$ paths for each UT.
We consider a ULA topology at the BS with $0.5 \lambda$ antenna spacing.
The channel is generated according to \eqref{eq:1}, where the AoDs and AoAs are randomly distributed.
\begin{figure}[!t]
\centering
{\includegraphics[width=0.45\textwidth]{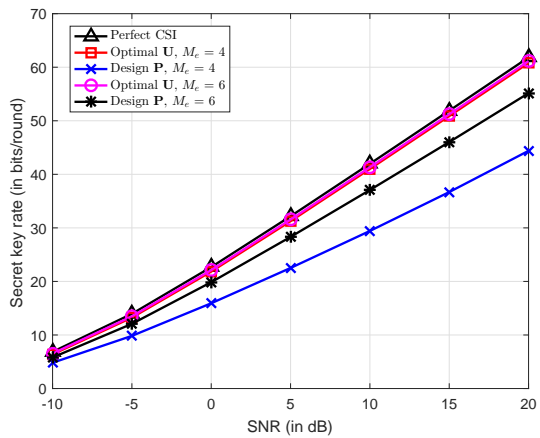}}
\caption{Secret key rate comparison for one UT.}\label{fig:F1}
\end{figure}

Fig.~\ref{fig:F1} presents the secret key rate of single user to confirm that our proposed CDR based secret key generation scheme is also suitable for the single user case.
We compare the secret key rate of the designed matrices $\U$ and $\tilde{\P}$ with that of perfect CSI. The perfect CSI provides the complete channel information and achieves the highest secret key rate. Here, we set the matrix $\Ct = \I$ and consider $M_e=4$ and $M_e=6$ cases. From the results, when $M_e=6$, the secret key rate of optimal $\U$ and the designed $\Pt$ can approach that of perfect CSI, indicating that by employing the precoding matrix $\Pt$, the BS and the UT can obtain the almost perfect channel information, significantly reducing the dimension of the channel estimation and the pilot overhead.
When $M_e = 4$, the secret key rate approaches that of $M_e = 6$, which contains the most channel power with lower overhead.

\begin{figure}[!t]
\centering
{\includegraphics[width=0.45\textwidth]{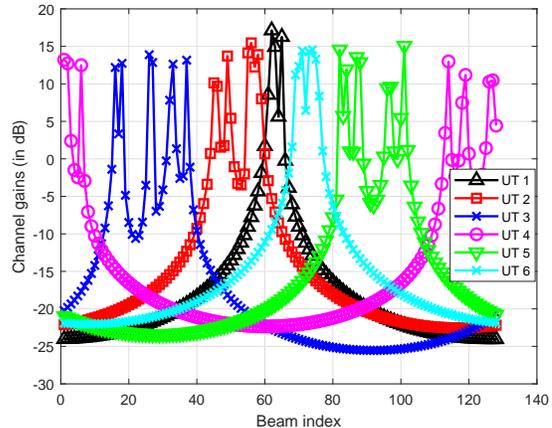}}
\caption{Multi-user channel gains distribution in the beam domain. }
\label{fig:Beam4}
\end{figure}

Next, we consider the multi-user secret key generation and illustrate an example of multi-user channel gains distribution in the beam domain in Fig.~\ref{fig:Beam4}.
{The BS employs the eigenmatrix of the channel $\A_{BS}$ to generate $M$ fixed beams of different directions, where the $m$-th beam is corresponding to the direction $\sin(\varphi_m) = 2m/M-1$.
Then, according to the particular location of the UT, the BS selects a number of beams from the $M$ beams to generate secret key with him.
}

When six UTs are distributed in different positions, the channel gains of each UT are concentrated within a number of beams (or directions), different UTs occupy non-overlapping channel beams. The attenuation between the adjacent UTs is about 20~dB, significantly reducing inter-user interference.
This result indicates that the BS equipped with massive antennas has the potential to achieve multi-user secret key generation.

\begin{figure}[!t]
\centering
{\includegraphics[width=0.45\textwidth]{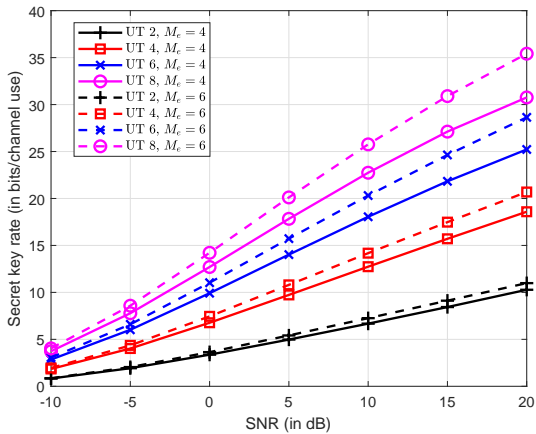}}
\caption{Secret key rate comparison for different number of UTs.}\label{fig:UT}
\end{figure}

{
Fig.~\ref{fig:UT} compares the secret key rate for different number of users. From the results, we can find that as the number of UTs increases, the secret key rate grows up approximately linearly. For example, the key rate of 4 UTs approaches twice than that of 2 UTs. However, as the number of UTs continues increasing, the key rate of each UT becomes lower, due to the interference or information leakage among UTs.
}

\begin{figure}[!t]
\centering
{\includegraphics[width=0.45\textwidth]{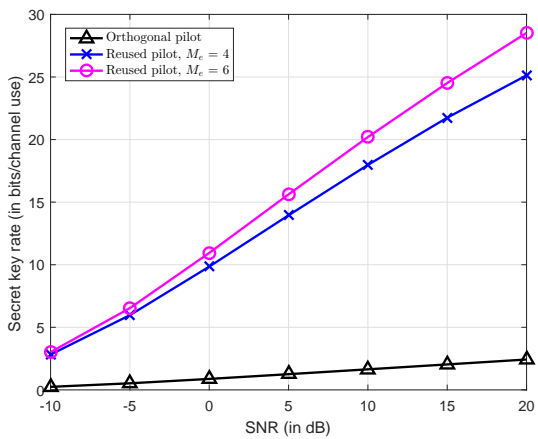}}
\caption{Unit secret key rate comparison for multiple UTs of orthogonal pilot and reused pilot.}\label{fig:F3}
\end{figure}

In multi-user secret key generation, the bottleneck is the pilot overhead.  Considering the negative effect of pilot overhead, we define the unit secret key rate as
\begin{align}
	R_{\text{unit}} = R_{\text{sum}}/T,
\end{align}
where $T$ is the pilot overhead, scaled with the dimension of the effective channel $M_e$ and $N_e$.
As the number of antennas at each UT is $4$, we set $N_e=N_k=4$.
Fig.~\ref{fig:F3} compares the unit secret key rate of reused pilot with $M_e=4$ and $M_e=6$ with orthogonal pilot scheme. As the pilot overhead is extremely large for orthogonal pilot scheme, the unit secret key rate suffers serious loss. The reused pilot scheme with $M_e=6$ achieves the highest rate and the rate of scheme with $M_e=4$ is close to that of $M_e=6$.

\begin{figure}[!t]
\centering
{\includegraphics[width=0.45\textwidth]{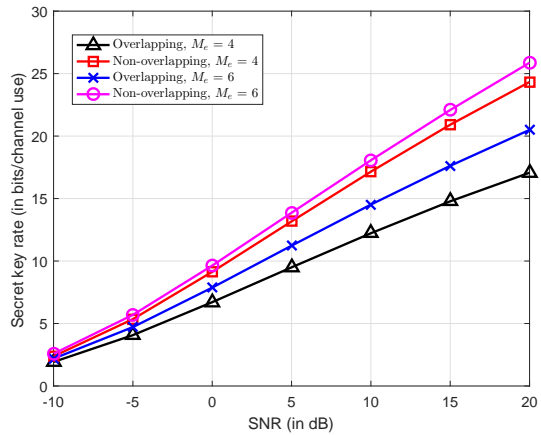}}
\caption{Unit secret key rate comparison of overlapping and non-overlapping beams.}\label{fig:F4}
\end{figure}

\begin{figure}[!t]
\centering
{\includegraphics[width=0.45\textwidth]{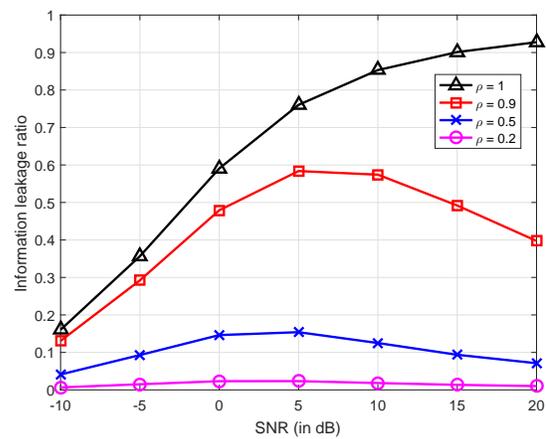}}
\caption{Information leakage ratio with different correlation coefficients.}
\label{fig:}
\end{figure}

Fig.~\ref{fig:F4} compares the unit secret key rate of overlapping and non-overlapping transmitting beam schemes, when the channel beams are overlapping between different UTs. For the overlapping scheme, the BS allocates the strongest transmitting beams for each UT, some of which may be overlapping with other UTs, while for the non-overlapping scheme, the BS allocates the non-overlapping strongest transmitting beams for each UT.
For the overlapping transmitting beam scheme, to estimate the channel of overlapping beams for different users, orthogonal pilot is used. Thus, the overhead is a little larger than that of non-overlapping transmitting beam scheme. Here, we do not consider the information leakage of the overlapping beams and only consider the interference between users. We observe that the unit secret key rates of non-overlapping schemes are higher than that of overlapping schemes. The non-overlapping scheme with $M_e=6$ achieves the highest rate.

Then, we present the information leakage ratio when the channels of different UTs are correlated, as shown in Fig.~\ref{fig:}.
{If the channels of UT $1$ and UT $2$ are correlated, they may observe similar channel measurements, resulting in the information leakage. When UT $2$ is a potential eavesdropper, it can guess part of the key of UT $1$, according to its correlated channel measurement. }
From the result, when the correlation coefficient is $1$, the information leakage ratio increases as the SNR grows up. When the correlation coefficient is less than 1, the information leakage ratio increases in the low and middle SNR regions and decreases in the high SNR regions. This is because in the high SNR regions, the BS can obtain the precise channel information and extract the difference between them to generate the keys.
However, the information leakage is still large when the correlation is high.

\begin{figure}[!t]
\centering
{\includegraphics[width=0.45\textwidth]{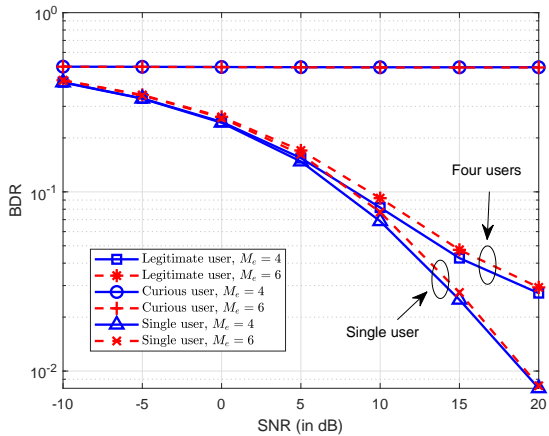}}
\caption{BDR comparison for legitimate and curious UTs.}
\label{fig:BDR}
\end{figure}

{
Next, we evaluate the bit disagreement ratios (BDR) performance of legitimate user and curious user, as shown in Fig.~\ref{fig:BDR}.
The BDR is defined as the ratio of the number of the disagreement bits to the number of total bits of the initial secret key, which is the quantization result of channel measurement.
In the figure, the curve of ``legitimate user'' refers to the BDR between the BS and each UT, while the curve of ``curious user'' refers to the BDR of two different UTs,
which presents the key disagreement between different UTs.
Moreover, we also illustrate the BDR of single user case, which is the best case without inter-user interference.
From the results, we can find that the BDR of  ``legitimate user'' approaches that of
``single user'', which indicates that the BDR performance of our proposed multi-user
secret key generation approaches that of single-user key generation.
The BDR of ``curious user'' remains high (about $0.5$) for varied SNRs,
which means that the quantized channel measurements of different UTs are different,
one UT cannot guess the key of other UT based on his observation.

Finally, we evaluate the randoness of the initial key (before privacy amplification) via the National Institute of Standards and Technology (NIST) random test suite~\cite{randomnessteset}.
A tested bit sequence passes a test when the p-value is greater than the threshold, usually
chosen as $0.01$. We perform 9 NIST statistical tests for $10000$ trials, and each initial key has a length of $256$ bits.
The pass ratios and the averaged p-values are summarized in Table \ref{table1}.
For each test, the pass ratio is higher than $90\%$ and the averaged p-value is significantly greater than $0.01$. The results reflect a good randomness of the initial key generated via our proposed approach.

\begin{table}[htbp]
    \centering
    \caption{NIST random test result.}
    \vspace{1em}
    \label{table1}
    \begin{tabular}{|c|c|c|c|c|c|c|c|c|c|}
        \hline
        & Pass ratio & P-value\\
        \hline
        Approximate entropy & 0.9199 & 0.4233 \\ \hline
        Runs &  0.9262 & 0.4590 \\ \hline
        Ranking &  0.9128 & 0.3810 \\ \hline
        Longest runs of ones & 0.9717 & 0.3701 \\ \hline
        Frequency & 0.9926 & 0.5254 \\ \hline
        FFT & 0.9983 & 0.5800 \\ \hline
        Block frequency & 0.9949  & 0.5466 \\ \hline
        Cumulative sums & 0.9974 & 0.4997\\ \hline
        Serial & 0.9255 & 0.4474, 0.4903\\
        \hline
    \end{tabular}
\end{table}

}

\section{Conclusion} 
\label{sec:conclusion}
This paper provided a fundamental design and analysis of the multi-user secret key generation in massive MIMO wireless networks. We provided a beam domain channel model, representing the channel gains from different transmit directions to different receive directions. We derived a closed-form expression of the secret key rate, which depends on the statistical CSI and the precoding and receiving matrices. We provided the optimal conditions for the Kronecker of the precoding the receiving matrices and proposed an algorithm to achieve the maximal {sum} secret key rate. When the beams of different UTs are non-overlapping, the BS employs several strongest beams of each UT to simultaneously generate secret key. Furthermore, we provided a security analysis by considering the channel correlation between UTs. When the channels of different UTs are correlated, the BS employs the several strongest non-overlapping beams of each UT to generate secret key.
Numerical results demonstrate the performance improvement of our proposed multi-user secret key generation scheme.
{This work focuses on the sum secret key rate maximization, while the power allocation optimization under the fairness constraint among UTs can be further analyzed in the future.}

\appendices
\section{Proof of Proposition \ref{prop:1}} 
\label{sec:proof_of_proposition_prop:1}

From \eqref{eq:6}, the $(n,m)$th element of the beam domain channel $\Ht_{k}^{DL}$
can be expressed as
\begin{align}
    &[\Ht_{k}^{DL}]_{n,m} = \a_{UT,k}(\theta_n)^H \H_{k}^{DL} \a_{Bs}(\varphi_m) \nn\\
    &= \sum_p \alpha_{k,p} \a_{UT,k}(\theta_n)^H  \a_{UT,k}(\theta_{k,p})  \a_{BS}(\varphi_{k,p})^H   \a_{BS}(\varphi_m).
\end{align}
First, we consider the calculation of $\a_{UT,k}(\theta_n)^H  \a_{UT,k}(\theta_{k,p})$.
As the number of UT antennas tends to infinity,
there exists $\theta_n$ equal to $\theta_{k,p}$ ($\theta_n = \theta_{k,p}$),
and
\begin{align}
    \a_{UT,k}(\theta_n)^H  \a_{UT,k}(\theta_{k,p}) = 1.
\end{align}
When $\theta_n$ is not equal to $\theta_{k,p}$,
we have \cite{7227112}
\begin{align}
    &\lim_{N\to\infty }\a_{UT,k}(\theta_n)^H  \a_{UT,k}(\theta_{k,p})\nn\\& = \lim_{N_k\to\infty }\frac{1}{N_k} \frac{1- e^{-j\frac{2\pi}{\lambda}d N_k (\sin(\theta_{k,p})-\sin(\theta_n))}}{{1- e^{-j\frac{2\pi}{\lambda}d(\sin(\theta_{k,p})-\sin(\theta_n))}}} =0.
\end{align}
Similarly, as the number of BS antennas grows,
we have
\begin{align}
    \lim_{M\to\infty} \a_{BS}(\varphi_{k,p})^H   \a_{BS}(\varphi_m) = \delta(\varphi_{k,p}- \varphi_{m}).
\end{align}
Thus, the $(n,m)$th element of $\Ht_{k,p}$ can be expressed as
\begin{align}
    \lim_{N,M\to\infty} [\Ht_{k}^{DL}]_{n,m} -\sum_p \alpha_{k,p} \delta(\theta_k,p-\theta_n)\delta(\varphi_{k,p}- \varphi_{m}) = 0.
\end{align}
This completes the proof.\qed

\section{Proof of Theorem \ref{thm:1}} 
\label{sec:proof_of_theorem_thm:1}

Assuming zero-mean complex Gaussian random vector for each channel observation $\z_k^{DL}$ or $\z_k^{UL}$, we have \cite{5483148}
\begin{align}\label{eq:I1}
	I (\z_k^{DL};\z_k^{UL} | \z_i^{DL}  )
	&= H(\z_k^{DL}, \z_i^{DL}) + H(\z_k^{UL} , \z_i^{DL} )\nn\\
    &\quad- H(\z_k^{DL}, \z_k^{UL} , \z_i^{DL}) - H(\z_i^{DL}) \nn\\
	&= \log \frac{\det( \Rc_{\z_k^{DL} \z_i^{DL}} \Rc_{\z_k^{UL} \z_i^{DL}} )}{\det( \Rc_{\z_k^{DL} \z_k^{UL} \z_i^{DL}}) \det(\Rc_{ \z_i^{DL}} )}.
\end{align}
Specially, when the channel observations of different UTs are uncorrelated,
i.e., the channels of different UTs are independent,
the conditional mutual information \eqref{eq:I1} can be simplified as
\begin{align}
	I (\z_k^{DL};\z_k^{UL} | \z_i^{DL}  )  = I (\z_k^{DL};\z_k^{UL}  )
	=  \log \frac{\det( \Rc_{\z_k^{DL} } \Rc_{\z_k^{UL}} )}{\det( \Rc_{\z_k^{DL} \z_k^{UL} } )},
\end{align}
which only depends on the correlation of uplink and downlink channels.

We will calculate the covariance matrices $\Rc_{\z_k^{DL} }$, $\Rc_{\z_k^{UL}}$, and $\Rc_{\z_k^{DL} \z_k^{UL} }$ to obtain the secret key rate.
The matrix $\Rc_{ \z_k^{DL}} $ can be calculated as
\begin{align}
    \Rc_{ \z_k^{DL}} =& \Eb \Bigg\{  \sum_{k'}((\P_{k'}\S^{DL}_{k'k})^T \otimes \C_k^H) \tvec(\H_k) \tvec(\H_k)^H \nn\\
    &\times \sum_{k'}((\P_{k'}\S^{DL}_{k'k})^T \otimes \C_k^H)^H \nn\\
    & + (\I_{T_D} \otimes \C_k^H )\tvec(\N_k)\tvec(\N_k)^H(\I_{T_D} \otimes \C_k^H )^H     \Bigg\}.
\end{align}
Without loss of generality, we consider the unit covariance matrix of noise, given by
\begin{align}
    \Eb \bigg\{ \tvec(\N_k)\tvec(\N_k)^H  \bigg\} = \I_{NM_e}.
\end{align}
Recalling $\R_k = (\A_{BS}^* \otimes \A_{UT}) \Lambdam_{k} (\A_{BS}^* \otimes \A_{UT})^H$,
we have
\begin{align}
    \Rc_{ \z_k^{DL}} =&  \sum_{k'}((\P_{k'}\S^{DL}_{k'k})^T \otimes \C_k^H)\R_k \sum_{k'}((\P_{k'}\S^{DL}_{k'k})^T \otimes \C_k^H)^H\nn\\& + (\I_{T_D} \otimes \C_k^H )(\I_{T_D} \otimes \C_k^H )^H \nn\\
    =& \sum_{k'}((\A_{BS}^H\P_{k'}\S^{DL}_{k'k})^T \otimes \C_k^H\A_{UT}) \Lambdam_{k}\nn\\ & \times\sum_{k'}((\A_{BS}^H\P_{k'}\S^{DL}_{k'k})^T \otimes \C_k^H\A_{UT})^H \nn\\ &+ (\I_{T_D} \otimes \C_k^H \C_k).
\end{align}
Let $\Pt_k = \A_{BS}^H\P_{k}$ and $\Ct_k = \A_{UT,k}^H\C_k$.
The covariance matrix $\Rc_{ \z_k^{DL}}$ can be rewritten as
\begin{align}
    \Rc_{ \z_k^{DL}} =& \Big(\sum_{k'}(\Pt_{k'}\S^{DL}_{k'k})^T \otimes \Ct_k^H\Big) \Lambdam_{k} \Big(\sum_{k'}(\Pt_{k'}\S^{DL}_{k'k})^T \otimes \Ct_k^H\Big)^H
    \nn\\&+ (\I_{T_D} \otimes \C_k^H \C_k).
\end{align}
Similarly, we can calculate $\R_{\z_k^{UL}}$ as
\begin{align}
    \Rc_{ \z_k^{UL}} =&\sum_{k'}\Big(\Pt^T_k\otimes \S^{UL}_{k'k}\Ct^H_{k'}\Big)
     \Lambdam_{k'} \Big(\Pt^T_k \otimes \S^{UL}_{k'k}\Ct^H_{k'}\Big) ^H \nn\\&+ (\P^T_k\P^*_k \otimes \I_{T_U}).
\end{align}

Next, we will calculate the determinant of covariance matrix $\Rc_{\z_k^{DL}\z^{UL}_k}$. Note that the matrix $\Rc_{\z_k^{DL}\z^{UL}_k}$ can be decomposed as
\begin{align}
    \Rc_{\z_k^{DL}\z^{UL}_k} = \begin{bmatrix}
        \Rc_{\z_k^{DL}} & \R_{\z_k^{DL} \z_k^{UL}}  \\
        \R_{\z_k^{UL} \z_k^{DL}}  &  \Rc_{\z_k^{UL}}
    \end{bmatrix}
\end{align}
where $\R_{\z_k^{DL} \z_k^{UL}}$ represents the covariance of $\z_k^{DL}$ and $\z_k^{UL}$,
\begin{align}
    \R_{\z_k^{DL} \z_k^{UL}} =& \Eb \{ \z_k^{DL} (\z_k^{UL})^H\} \nn\\
    =&\Big(\sum_{k'}(\Pt_{k'}\S^{DL}_{k'k})^T \otimes \Ct_k^H\Big) \Lambdam_k
     \Big(\Pt^T_k\otimes \S^{UL}_{kk}\Ct^H_{k}\Big)  ^H .
\end{align}
From the determinant of the block matrix, we have
\begin{multline}
    \det(\Rc_{\z_k^{DL}\z^{UL}_k})\\ = \det(\Rc_{\z_k^{DL}})
    \det\Big(\Rc_{\z_k^{UL}} - \R_{\z_k^{UL} \z_k^{DL}}\Rc_{\z_k^{DL}}^{-1}\R_{\z_k^{DL} \z_k^{UL}} \Big).
\end{multline}
Hence, the secret key rate can be expressed as
\begin{align}\label{eq:57}
    I(\z_k^{DL};\z_k^{UL}) &= \log \frac{\det (\Rc_{\z_k^{UL}})}{\det\Big(\Rc_{\z_k^{UL}} - \R_{\z_k^{UL} \z_k^{DL}}\Rc_{\z_k^{DL}}^{-1}\R_{\z_k^{DL} \z_k^{UL}} \Big)} \nn\\
    = -&\log \det \left( \I - \Rc_{\z_k^{UL}}^{-1}\R_{\z_k^{UL} \z_k^{DL}}\Rc_{\z_k^{DL}}^{-1}\R_{\z_k^{DL} \z_k^{UL}} \right).
\end{align}
Let $\V_k = \Lambdam_k^{1/2}\Big(\sum_{k'}(\Pt_{k'}\S^{DL}_{k'k})^T \otimes \Ct_k^H\Big)^H$ and $\V_{kk'} = \Lambdam_{k'}^{1/2} \Big(\Pt^T_k\otimes \S^{UL}_{k'k}\Ct^H_{k'}\Big)^H$.
Then, we can have
\begin{align}
    & \Rc_{\z_k^{UL}}^{-1}\R_{\z_k^{UL} \z_k^{DL}}\Rc_{\z_k^{DL}}^{-1}\R_{\z_k^{DL} \z_k^{UL}}\nn\\
    &= \left(\sum_{k'} \V_{kk'}^H \V_{kk'}  + (\P^T_k\P^*_k \otimes \I_{T_U})\right)^{-1}\nn\\
    &\times \V_{kk}^H \V_k \left(  \V_k^H\V_k + \I_{T_D} \otimes \C_k^H \C_k \right)^{-1}\V_k^H \V_{kk}
\end{align}
and the secret key rate is given by
\begin{align}
    &I(\z_k^{DL};\z_k^{UL}) \nn\\
    &= - \log \det \left(\I - \V_{kk}  \left(\sum_{k'} \V_{kk'}^H \V_{kk'}  + (\P^T_k\P^*_k \otimes \I_{T_U})\right)^{-1} \right.\nn\\
    &\left. \times \V_{kk}^H \V_k \left(  \V_k^H\V_k + \I_{T_D} \otimes \C_k^H \C_k \right)^{-1}\V_k^H \right).
\end{align}
This completes the proof. \qed

\section{Proof of Theorem \ref{thm:2}} 
\label{sec:proof_of_proposition_thm:1}
The sum rate can be expressed as
\begin{align}
    R_k \!=\! -\!\log\!\det\!\left(\! \I\! - \!\left(\! \Lambdam_k^{1/2} \U_k \! \left(\I + \U_k^H \Lambdam_{k} \U_k \right)^{-1}\!\U_k^H  \Lambdam_k^{1/2}\!\right)^2 \right).
\end{align}
From Eq. 10.55 in \cite{seber2008matrix}, we have
\begin{align}
    &\Lambdam_k^{1/2} \U_k (\I + \U_k^H \Lambdam_k \U_k)^{-1}\U_k^H \Lambdam_k^{1/2} \nn\\
    &= \Lambdam_k^{1/2} \U_k \Big(\U_k^H (\I +  \Lambdam_k) \U_k\Big)^{-1}\U_k^H \Lambdam_k^{1/2} \nn\\
    & \preceq \Lambdam_k^{1/2}  (\I +  \Lambdam_k)^{-1}  \Lambdam_k^{1/2}.
\end{align}
Thus, the sorted eigenvalues satisfy
\begin{multline}
    \lambda_i (\Lambdam_k^{1/2} \U_k (\I + \U_k^H \Lambdam_k \U_k)^{-1}\U_k^H \Lambdam_k^{1/2} ) \\
    \le \lambda_i (\Lambdam_k (\I +  \Lambdam_k)^{-1}  )
\end{multline}
and we have
\begin{align}
    \lambda_i \Big(
    \I -  \left( \Lambdam_k^{1/2} \U_k  \left(\I + \U_k^H \Lambdam_{k} \U_k \right)^{-1}\U_k^H  \Lambdam_k^{1/2}\right)^2
    \Big)  \nn\\ \ge \lambda_i \Big(\I - \Lambdam_k^{2}  (\I +  \Lambdam_k)^{-2}  \Big).
\end{align}
Thus,
\begin{align}
    R_k &= - \sum_i \log \lambda_i \Big(
    \I -  \left( \Lambdam_k\U_k  \left(\I + \U_k^H \Lambdam_{k} \U_k \right)^{-1}\U_k^H  \right)^2
    \Big) \nn\\
    &\le -\sum_i \log \lambda_i \Big(\I - \Lambdam_k^{2}  (\I +  \Lambdam_k)^{-2}  \Big).
\end{align}
The equality holds only when $\U_k$ is consist of the unit vectors with the indices of unit elements corresponding to that of the sorted eigenvalues, i.e.,
\begin{align}
    \U_k = \begin{bmatrix}
        \e_{\lambda_1} & \e_{\lambda_2} & \cdots & \e_{\lambda_{M_eN_e}}
    \end{bmatrix}
\end{align}
where $\e_i = [0,0,\cdots,0,1,0,\cdots,0]$ is a unit vector with the $i$th unit element.
This completes the proof. \qed

\section{Proof of Theorem \ref{thm:3}} 
\label{sec:proof_of_proposition_thm:3}

The secret key rate of UT $1$ corresponding to beam $b$ is given by
\begin{small}
    \begin{align}
    R_{1}& = \log \frac{\det\left(\Rc_{(h_1+n_1^{DL})(h_2+n_2^{DL})}\Rc_{(h_1+n_1^{UL})(h_2+n_2^{DL})}\right)}{\det\left(\Rc_{(h_1+n_1^{DL})(h_1+n_1^{UL})(h_2+n_2^{DL})} \right)\det\left(  \Rc_{(h_2+n_2^{DL})}\right)}
\end{align}
\end{small}
where $n_1^{UL}$ and $n_1^{DL}$ (or $n_2^{DL}$) are the received noise of UT $1$ (or UT $2$) in the uplink and downlink transmissions.
When the variance of the noise is $\sigma^2$, we can calculate $R_{b1}$ in closed-form as
\begin{align}
    R_{1}  = \log\frac{((1+\sigma^2)^2-\rho^2)^2}
    {(1+\sigma^2)(\sigma^6+3 \sigma^4-2 \sigma^2\rho^2 + 2 \sigma^2)}.
\end{align}

Taking the derivative of $R_{b1}$ with respect to $\rho$, we can obtain
\begin{align}
    \frac{\partial R_{b1}}{\partial \rho} = \frac{-4 \rho\sigma^2 (1 + \sigma^2 - \rho)  (1 + \sigma^2 + \rho) (1 + \sigma^2 - \rho^2)}{(1 + \sigma^2) (2 \sigma^2 + 3 \sigma^4 + \sigma^6 - 2 \sigma^2 \rho^2)^2}.
\end{align}
As $\rho$ is in the region $[-1,1]$, $(1 + \sigma^2 - \rho)$, $(1 + \sigma^2 + \rho)$, and $(1 + \sigma^2 - \rho^2)$ in the numerator are positive.
Then, when $\rho$ is in the region $[0,1]$, the derivative $\frac{\partial R_{b1}}{\partial \rho}\le 0$,
which indicates the rate $R_{b1}$ is monotonic decreasing.
When $\rho$ is in the region $[-1,0]$, the derivation $\frac{\partial R_{b1}}{\partial \rho}\ge 0$,
which indicates the rate $R_{b1}$ is monotonic increasing.
Thus, we can have the highest rate $R_{h}$ with $\rho=0$ as
\begin{align}
     R_{h} = \log\frac{(1 + \sigma^2)^2}
    {\sigma^2 (2 + \sigma^2)},
\end{align}
and the lowest rate $R_{l}$ with $\rho=1$ or $\rho=-1$ as
\begin{align}\label{eq:56}
    R_{l} = \log \frac{(2 + \sigma^2)^2}{3 + 4\sigma^2 + \sigma^4}.
\end{align}
Thus, the information leakage ratio can be calculated as
\begin{align}
	\gamma = 1- \frac{\log\frac{((1+\sigma^2)^2-\rho^2)^2}
    {(1+\sigma^2)(\sigma^6+3 \sigma^4-2 \sigma^2\rho^2 + 2 \sigma^2)}}
	{\log\frac{(1 + \sigma^2)^2}{\sigma^2 (2 + \sigma^2)}}.
\end{align}
This completes the proof. \qed

\section*{Acknowledgment}
This work of G. Li is partly funded by the National Natural Science Foundation of China under Grant 61801115 and 61941115, in part by the National key research and development program of China under Grant 2020YFE0200600, in part by the Zhishan Youth Scholar Program of SEU under Grant 3209012002A3.
The work of C. Sun is partly funded by the National Natural Science Foundation
of China under Grants 61901110, the Natural Science Foundation of Jiangsu Province under Grant BK20190334.
The work of E. Jorswieck is partly funded by the German Research Foundation (DFG) under project JO 801/25-1.

\bibliographystyle{IEEEtran}
\bibliography{IEEEabrv,mybibfile}

%

\end{document}